\documentclass[lettersize,journal]{IEEEtran}
\usepackage{amsmath,amsfonts}
\usepackage{algorithmic}
\usepackage{algorithm}
\usepackage{array}
\usepackage[caption=false,font=normalsize,labelfont=sf,textfont=sf]{subfig}
\usepackage{textcomp}
\usepackage{stfloats}
\usepackage{url}
\usepackage{verbatim}
\usepackage{graphicx}
\usepackage{cite}
\usepackage{color}
\usepackage{amssymb}
\usepackage{multirow}
\usepackage{flushend}
\usepackage[backref]{hyperref}

\newcommand{\rmD}{\mathrm{D}}

\newcommand{\rmX}{\mathrm{X}}

\newcommand{\rmG}{\mathrm{G}}
\newcommand{\rmL}{\mathrm{L}}
\newcommand{\rmS}{\mathrm{S}}
\newcommand{\rmF}{\mathrm{F}}
\newcommand{\rmA}{\mathrm{A}}
\newcommand{\rmV}{\mathrm{V}}

\newcommand{\ba}{\mathbf{a}}

\newcommand{\bY}{\mathbf{Y}}

\newcommand{\bS}{\mathbf{S}}

\newcommand{\bN}{\mathbf{N}}

\newcommand{\bH}{\mathbf{H}}
\newcommand{\bU}{\mathbf{U}}

\newcommand{\bp}{\mathbf{p}}

\newcommand{\CN}{\mathcal{CN}}

\newcommand{\bbC}{\mathbb{C}}

\newcommand{\calN}{\mathcal{N}}

\newcommand{\ctrans}{\mathsf{H}}

\newcommand{\LoS}{\mathrm{LoS}}

\begin{document}

\title{Distributed U6G ELAA Communication Systems: \\Channel Measurement and Small-Scale \\ Fading Characterization}

\author{
Jiachen Tian,~\IEEEmembership{Graduate Student Member,~IEEE,}
Zhengtao Jin,
Xiayang Chen,
Yu Han,~\IEEEmembership{Member,~IEEE,}\\
Shi Jin,~\IEEEmembership{Fellow,~IEEE,}
Wenjin Wang,~\IEEEmembership{Member,~IEEE}, and
Chao-Kai Wen,~\IEEEmembership{Fellow,~IEEE}

\thanks{J. Tian, Z. Jin, Y. Han, S. Jin and W. Wang are with the National Mobile Communication Research Laboratory, Southeast University, Nanjing 210096, China (email: \{tianjiachen, jinzhengtao, hanyu, jinshi, wangwj\}@seu.edu.cn).}
\thanks{X. Chen is with the Chien-Shiung Wu College, Southeast University, Nanjing 210096, China (email: 213221682@seu.edu.cn)}.
\thanks{C.-K. Wen is with the Institute of Communications Engineering, National Sun Yat-sen University, Kaohsiung 80424, Taiwan (e-mail: chaokai.wen@mail.nsysu.edu.tw).}
}



\maketitle

\begin{abstract}
The distributed upper 6 GHz (U6G) extra-large scale antenna array (ELAA) is a key enabler for future wireless communication systems, offering higher throughput and wider coverage, similar to existing ELAA systems, while effectively mitigating unaffordable complexity and hardware overhead. Uncertain channel characteristics, however, present significant bottleneck problems that hinder the hardware structure and algorithm design of the distributed U6G ELAA system. In response, we construct a U6G channel sounder and carry out extensive measurement campaigns across various typical scenarios. Initially, U6G channel characteristics, particularly small-scale fading characteristics, are unveiled and compared across different scenarios. Subsequently, the U6G ELAA channel characteristics are analyzed using a virtual array comprising 64 elements. Furthermore, inspired by the potential for distributed processing, we investigate U6G ELAA channel characteristics from the perspectives of subarrays and sub-bands, including subarray-wise nonstationarities, consistencies, far-field approximations, and sub-band characteristics. Through a combination of analysis and measurement validation, several insights and benefits, particularly suitable for distributed processing in U6G ELAA systems, are revealed, which provides practical validation for the deployment of U6G ELAA systems.

\end{abstract}


\section{Introduction}
 
\IEEEPARstart{6G}{}wireless communication, the next evolutionary step in cellular wireless networks, is currently under intensive research but is facing the fundamental limitation of spectrum resources, which eventually hinders the pursuit of higher throughput \cite{6G}. Considering the spectral limitation of Sub-6 GHz bands and the severe transmission loss of millimeter-wave (mmW) frequency bands, academia and industry are beginning to pay attention to more promising frequency bands. The World Radiocommunication Conference 2023 (WRC-23) issued Resolution 245, which suggests that International Mobile Telecommunications (IMT) can potentially identify the frequency band upper 6 GHz (U6G), i.e., 6245–7125 MHz globally \cite{WRC-23}. Generally, 6G wireless communication systems are expected to operate in the 7–24 GHz band, known as frequency range 3 (FR3) \cite{38820}, which is expected to be integrated with existing wireless networks and realize consecutive microwave connectivity, i.e., Sub-6 GHz, 7–24 GHz, and mmWave \cite{PZZhang}.

In addition to spectrum resources, multi-antenna technologies are also an effective boost for transmission throughput. The extra large-scale massive multiple-input multiple-output (MIMO) is considered an augmented version of massive MIMO with an extended array size \cite{JYZhang,HQLu}. From another vantage point, the utilization of U6G is beneficial for integrating more antennas on the same array size and forming an extremely large-scale antenna array (ELAA) compared with the current Sub-6 GHz. Therefore, the integration of U6G and ELAA is considered a trend for future wireless communication systems \cite{WFan}, expected to enhance spectral efficiency (SE) and enable access for a massive number of users.

In pursuit of the blueprint outlined above, the deployment of U6G ELAA systems faces challenges due to uncertain channel characteristics, as well as high complexity and hardware costs introduced by the U6G frequency band and ELAA, respectively. Therefore, it is necessary to explore the channel characteristics, based on which distributed processing structures and corresponding transmission strategies with low complexity and cost can be derived.

\subsection{State-of-the-art}
\label{sec:relatedworks}
As a compromise strategy between Sub-6 GHz and mmW bands, the channel characteristics of the U6G band are crucial for researchers. For this purpose, the authors in \cite{JHZhangJSAC} compared the channel characteristics under 3.3 GHz, 6.5 GHz, 15 GHz, and 28 GHz, unveiling the effectiveness of mid-band spectrum in network coverage and transmission reliability. In \cite{TSR}, detailed comparisons of indoor channel characteristics were presented between emerging frequency bands belonging to FR1 and FR3, respectively. Furthermore, the authors in \cite{DMC} conducted a measurement campaign of a massive MIMO system working on the 11 GHz frequency band, which validates the characteristics of dense multipath component (DMC). To explore the performance of wireless communication systems working on this emerging frequency band, researchers also began to turn to ray-tracing (RT) simulations. In \cite{PZZhang}, the authors compared the signal-to-interference ratio (SINR) performance at six typical frequency points among FR3, providing an intuitive illustration of the SINR distribution. Similarly, the signal-to-noise ratio (SNR) and capacity performances were compared under four frequency bands within FR3, i.e., 6 GHz, 12 GHz, 18 GHz, and 24 GHz, in \cite{RTNewYork}. The coexistence of IMT services and satellite networks was also analyzed. Comparisons of coverage and throughput under FR1, FR2, and FR3 deployed in Boston were carried out in \cite{RTBoston}, illustrating the superiority of FR3 from the perspective of bandwidth. Although RT simulation is an effective means, modeling the characteristics of the U6G channel still relies on practical measurements, which remain insufficient.

From another perspective, the utilization of ELAA inspires novel channel characteristics such as spatial non-stationarities \cite{nonstationary,HanIOTJ}, which means different portions of ELAA experience different channel propagation characteristics. Specifically, spatial non-stationarities may manifest as near-field propagation characteristics, i.e., unequal phase differences between adjacent array elements, or unequal channel power among different array elements. From a theoretical research perspective, different non-stationary channel models and corresponding channel state information (CSI) acquisition schemes have been considered for ELAA systems \cite{TianWCL}. Based on the virtual array mechanism \cite{JHZhangST}, the non-stationarities under the mmW channel were validated based on measurements under four mmW frequency bands \cite{HuangJSAC}. The authors in \cite{JHZhang} measured the near-field channel characteristics of ELAA systems, while the near-field propagation characteristics were fitted and modeled. However, it can be seen that channel characteristics of U6G ELAA systems have not been further investigated from the perspective of subarray, which is directly related to distributed processing.

Considering the practical deployment of U6G ELAA systems, due to the significant increase in array dimensions, hardware costs and computational complexity become unaffordable. Different efficient hardware structures such as hybrid beamforming (HBF) structures have been proposed for massive MIMO in the 5G era. The authors in \cite{phase_switch} compared the energy efficiency of different combinations of phase shifters and switches in an HBF structure. Combined with ELAA systems, approximations and performance bounds of uplink achievable SE are derived in \cite{XYangTVT}, and efficient transmission strategies are proposed accordingly. Moreover, to deal with the high complexity introduced by near-field propagation characteristics, it is efficient to divide the ELAA into subarrays and apply far-field propagation approximation to each subarray \cite{LLDaiarXiv,CHan}. However, the subarray-wise far-field approximation of ELAA still lacks sufficient experimental validation. Furthermore, the deployment of low-cost hardware structures and low-complexity transmission strategies, especially integrated with distributed processing, still calls for the support of channel characteristics derived from practical measurement results.

\subsection{Key Problems and Contributions}
Given the potential of U6G ELAA wireless communication systems, the progress of U6G ELAA systems from a theoretical concept to practical deployment still faces the following challenges: 

\noindent \textbf{P1:} 
The channel characteristics of the U6G frequency band, especially when incorporated with ELAA, are still uncertain and call for in-depth investigation, including measurement and modeling. 

\noindent \textbf{P2:} 
Considering the unaffordable hardware cost and complexity introduced by the ELAA, the U6G ELAA channel characteristics need further inspection from a distributed perspective, which provides measurement validation for structures and algorithms of distributed processing. 

To tackle the aforementioned problems, we first measure and model the small-scale fading characteristics of the U6G channel. Afterwards, the channel characteristics of U6G ELAA systems are investigated from a distributed processing perspective, paving the way for distributed processing structures and algorithms. The contributions are threefold: 

\begin{itemize}

\item {\bf A U6G ELAA channel measurement system and campaigns across diverse scenarios}. We have developed an advanced time-domain channel sounding system operating in the U6G frequency band, utilizing a software-defined radio (SDR) platform and Xilinx RFSoC for multi-channel signal acquisition. By integrating the virtual array mechanism, we obtain comprehensive U6G ELAA channel data. Our extensive measurement campaigns span various scenarios, including indoor halls, outdoor environments, and outdoor-to-indoor (O2I) transitions, ensuring a robust and versatile dataset.

\item  {\bf In-depth analysis, modeling, and comparisons of U6G channel characteristics}. Employing a superresolution estimator, Newtonized orthogonal matching pursuit (NOMP) \cite{NOMP}, we extract multipath component (MPC) parameters to derive and analyze propagation characteristics in delay and angular domains. Key metrics such as average power angular spectrum (APAS), root mean square (RMS) angular spread (AS), power delay profile (PDP), RMS delay spread (DS), Rician K factor, and sparsity are meticulously examined. These characteristics are fitted and modeled to provide references for widely adopted channel models. Furthermore, we compare these metrics across different scenarios to deliver a comprehensive characterization of U6G channel features.
 
\item {\bf Pioneering analysis of U6G ELAA channel characteristics through distributed processing}. Drawing inspiration from distributed processing, we delve into the U6G ELAA channel characteristics by examining variations along the extended array elements and expanded bandwidth. Our investigation uncovers distributed traits such as subarray-wise nonstationarities, consistencies, far-field approximation, and sub-band characteristics. Through meticulous analysis of the measurement results, we unveil significant insights and benefits of distributed processing for ELAA systems, paving the way for innovative structures and groundbreaking transmission strategies.

\end{itemize}

The remainder of this paper is organized as follows.
Section \ref{sec:measure_setup} describes the measurement equipment, environment, and data processing methods. In Section \ref{sec:measure_result} and Section \ref{sec:ELAA}, the measurement results and channel characteristics of the U6G frequency band and U6G ELAA systems are presented, respectively, whilst insights and benefits of distributed processing are illustrated. Section \ref{sec:conclusion} provides the conclusions.

{\it Notations}--Vectors and matrices are denoted by bold lowercase and uppercase letters, respectively. The superscripts $(\cdot)^{\top}$ and $(\cdot)^{\mathsf{H}}$ represent the transpose and conjugate transpose, respectively; $\delta(\cdot)$ is the Dirac function; $\| \cdot \|_1$ and $\| \cdot \| _{\rm F}$ indicate the $\ell_1$-norm and the Frobenius norm, respectively. The zeroth-order modified Bessel function of the first kind is denoted as $I_0(\cdot)$. A Gaussian distribution with mean $\mu$ and variance $\sigma^2$ is denoted as $\calN(\mu,\sigma^2)$, and a complex Gaussian distribution is written as $\mathcal{CN}(\mu,\sigma^2)$.

\section{Measurement Equipment, Environment, and Data Processing}
\label{sec:measure_setup}
In this section, we describe the conducted measurement campaigns. The U6G channel sounder is primarily based on SDR and Xilinx RFSoC-based multichannel data acquisition equipment. With the aid of a virtual array mechanism, the static U6G ELAA channel characteristics can be obtained. Channel measurement campaigns are then carried out in typical scenarios, and the NOMP estimator is adopted for MPC acquisition.

\begin{figure}[!t]
    \centering
    \includegraphics[scale=0.43]{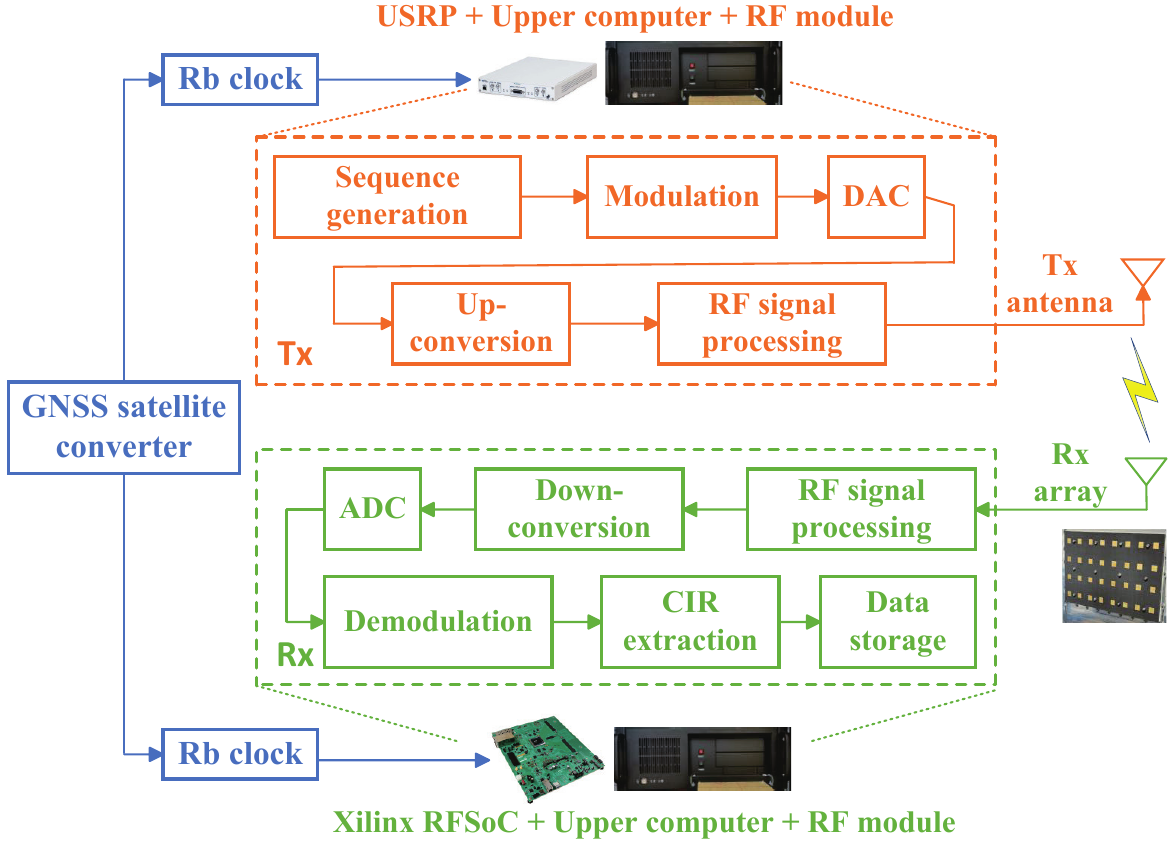}
    \caption{System setup of the U6G ELAA channel sounder.}
    \label{fig:system}
\end{figure}

\subsection{Measurement System Setup}

A time-domain channel measurement system is built to capture the characteristics of the U6G channel. As shown in Fig. \ref{fig:system}, the transmitter (Tx) of the system consists of an SDR universal software radio peripheral (USRP) 2943R, an upper computer, a radio frequency (RF) module, and a transmit antenna. The USRP internally contains baseband and RF signal processing modules, which are programmed by the LabVIEW software.

The detection signal is generated from the Zadoff-Chu (ZC) sequence and modulated into quadrature phase shift keying (QPSK) symbols. The symbols are then orthogonal frequency division multiplexing (OFDM) modulated with an effective bandwidth of 100 MHz and 3,276 effective subcarriers with a subcarrier space $\Delta f=30\ \text{kHz}$, while the sampling rate is 122.88 Msps and the number of fast Fourier transform (FFT) points is 4,096. The intermediate frequency (IF) detection signal is further up-converted to a 7.8 GHz central frequency for over-the-air (OTA) transmission.

The receiver (Rx) is composed of an antenna array, an RF module, multi-channel data acquisition equipment consisting of a Xilinx RFSoC, and an upper computer. The antenna array contains 32 elements, with 8 elements and 4 elements arranged in the horizontal and vertical directions, respectively. Two adjacent elements are spaced 2 cm apart, approximately half a wavelength. Every four elements in the vertical direction are combined and connected to the same digital channel. For each digital channel, the antenna gain is greater than or equal to 10 dBi, forming a narrow beam in the vertical direction and a wide beam in the horizontal direction. A uniform linear array (ULA) with a dimension of 8 is thus formed, and a virtual ELAA can be created by sliding the array horizontally.
 
Before measurement, the consistency calibration of multiple channels is carried out, and the equipment response is eliminated. In this paper, we focus only on the small-scale characteristics horizontally, and the Tx antenna and Rx antenna are placed at the same height with alignment between the Tx beam and Rx beams. During channel sounding, a single-stream signal is generated at the Tx and captured by multiple digital channels at the Rx. The equipment at Tx and Rx is accurately synchronized through a pair of GPS-disciplined Rubidium (Rb) clocks, with a 10 MHz square wave signal as a reference shared by all equipment.

\begin{figure*}[!t]
	\centering
    \subfloat[]{\includegraphics[width=.54\linewidth]{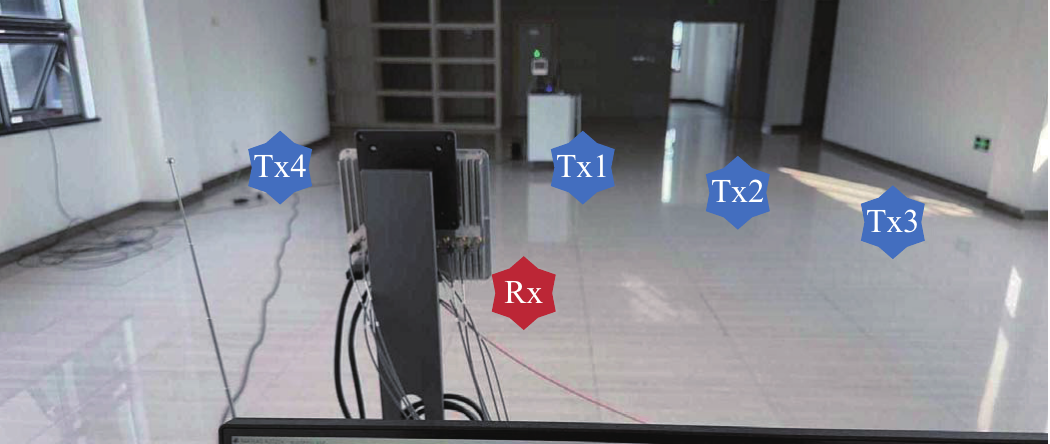}\label{fig:indoorhall}}\hspace{2pt}
	\subfloat[]{\includegraphics[width=.45\linewidth]{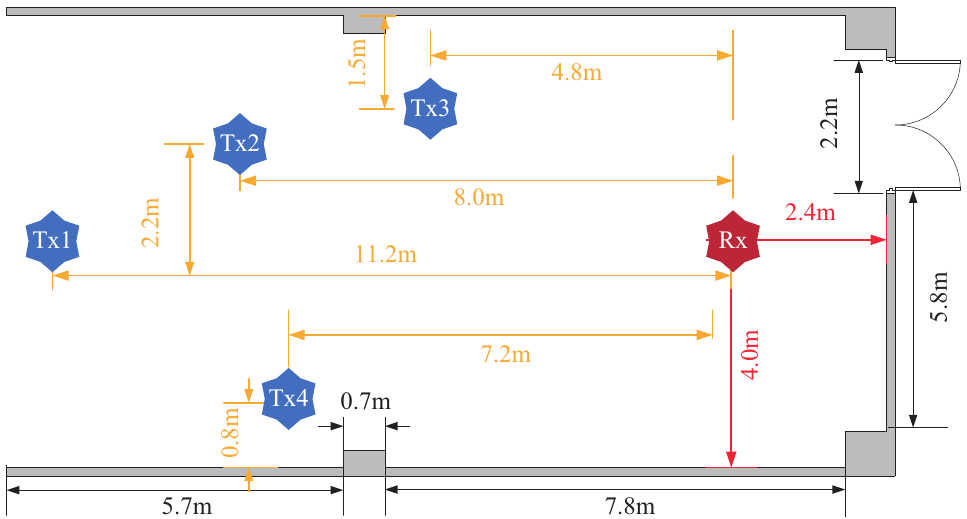}\label{fig:hall}}\\
    \subfloat[]{\includegraphics[width=.43\linewidth]{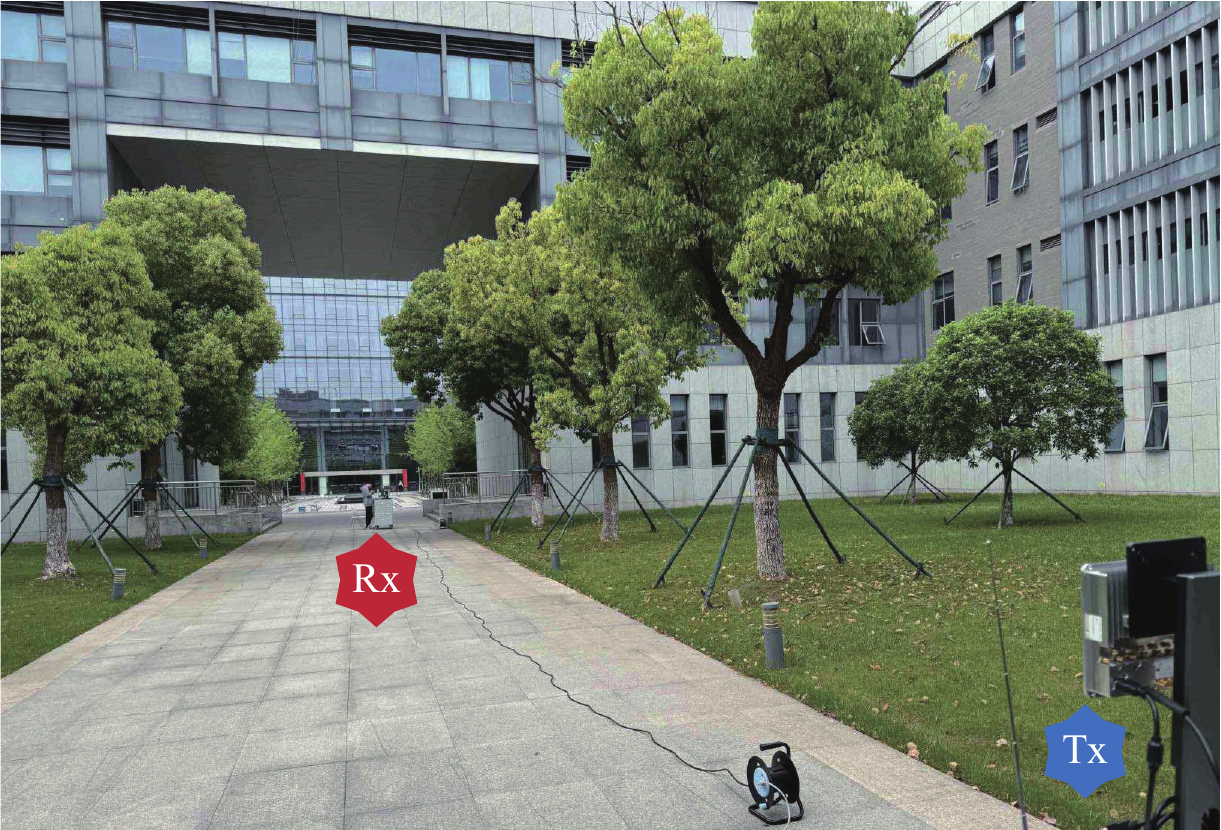}\label{fig:outdoor}}\hspace{1pt}
    \subfloat[]{\includegraphics[width=.29\linewidth]{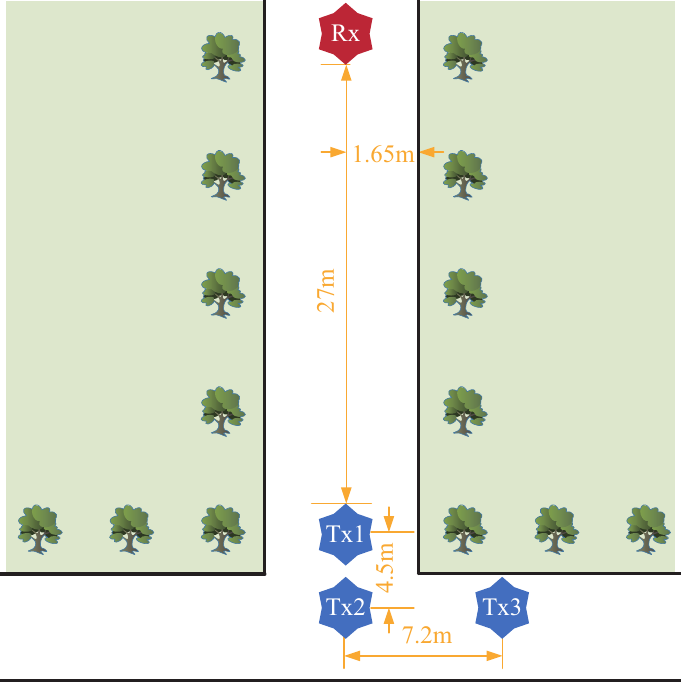}\label{fig:layout_outdoor}}\hspace{1pt}
	\subfloat[]{\includegraphics[width=.25\linewidth]{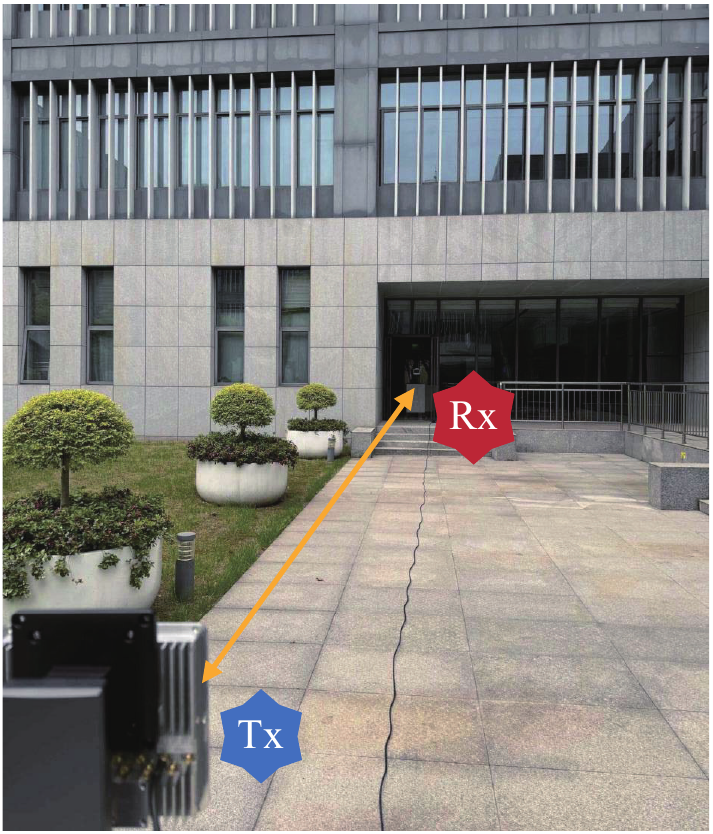}\label{fig:O2I}}\\
	\caption{Images of typical channel measurement campaigns: (a) Layout of indoor scenarios. (b) Schematic diagram of indoor scenarios. (c) Layout of outdoor scenarios. (d) Schematic diagram of outdoor scenario. (e) Layout of the O2I scenario.}
	\label{fig:scenarios}
\end{figure*}

\subsection{Measurement Environment}
The typical scenarios in these measurement campaigns are selected in the China Network Valley, Nanjing, China, including an indoor hall, outdoor, and O2I. During the measurements, all objects are kept stationary without human movement.

\subsubsection{Indoor Hall}
As shown in Fig. \ref{fig:scenarios}\subref{fig:indoorhall}, the indoor hall is approximately 14.2 m long and 8 m wide. Glass windows exist on the upper half of the two long sides, while the two short sides are concrete walls, and the floor is covered with smooth ceramic tiles. We selected four measurement points in total, and the positions are shown in Fig. \ref{fig:scenarios}\subref{fig:hall}, with different incident angles and distances between the Tx positions and Rx. We chose to fix the Rx in the center of the hall, 2.4 m away from the back wall. Tx is directly facing the Rx in Position 1, slightly towards the wall in Position 2, and very close to the wall in Positions 3 and 4. Note that the measurement positions are all under line-of-sight (LoS) conditions.

\subsubsection{Outdoor}
The outdoor measurement is carried out in an open area, and three Tx measurement positions are selected, as shown in Fig. \ref{fig:scenarios}\subref{fig:outdoor}.
A cement road surface exists between the Rx and the Tx, and the Rx is fixed at one end of the road. Sparse trees and grassy ground line both sides of the road. As shown in Fig. \ref{fig:scenarios}\subref{fig:layout_outdoor}, strong LoS links exist between Tx1 and Rx, and Tx2 and Rx, respectively, while the direct propagation link between Tx3 and Rx is slightly obstructed by trees, forming a weak LoS condition.

\subsubsection{O2I}
The layout of the O2I scenario is shown in Fig. \ref{fig:scenarios}\subref{fig:O2I}.
The Rx is located inside a building, and the Tx is on an open road. There are marble walls and glass doors near the Rx, and the Rx array is placed behind an open glass door. A LoS propagation link exists between the Tx and the Rx, and there are a few shrubs and grasslands on one side of the propagation path. The horizontal distance between the Tx and the Rx is about 19 m. The downtilt angles at Tx and Rx are adjusted to ensure beam alignment. 

\subsection{Data Processing}

After procedures including removing the pilot signal and cyclic prefix, the received OFDM signal in a symbol, $\bY$, is obtained.
Based on the widely-used ray-based channel model \cite{RBCM1,RBCM2}, the received signal is assumed to consist of a finite number of $L$ MPCs, i.e.,
\begin{equation}
    \bY = \sum_{\ell=1}^{L} \bS(\Theta _{\ell}) + \bN,
\end{equation}
where $\bS (\Theta _{\ell}) \in \bbC ^{N_{\rm f}\times N}$ represents the $\ell$-th MPC, $\Theta _{\ell}$ represents the parameters of the $\ell$-th MPC, $N_{\rm f}$ and $N$ are the number of subcarriers and antennas, respectively, $\bN$ is the additive random noise with elements satisfying $\CN(0,1)$. Note that when a ULA is deployed at the Rx, $\bS (\Theta _{\ell})$ can be further written as\footnote{In this paper, the MPC extraction is based on the small-scale array under the far-field assumption, while the MPCs of an ELAA are analyzed by dividing the ELAA into subarrays.}
\begin{equation}
    \bS(\Theta _{\ell}) = \alpha_{\ell} \bp(\tau _{\ell}) \ba ^{\ctrans}(\vartheta _{\ell}),
\end{equation}
where $\alpha _\ell$ is complex amplitude of the $\ell$-th MPC, $\bp(\tau _{\ell}) \triangleq [p _1(\tau _{\ell}),\dots,p_{N_{\rm f}}(\tau _{\ell})] ^{\top}$, with $p _{n_{\rm f}}(\tau _{\ell}) = \exp(-\jmath 2 \pi \eta_{\rm f} \Delta f \tau _\ell/\lambda)$, $\eta _{\rm f}$ is the index of subcarrier, $\ba(\vartheta_\ell) \triangleq [ a_1(\vartheta_{\ell}), \dots, a_N (\vartheta _{\ell})] ^{\top}$ is the steering vector of the array, with $a_n(\vartheta _{\ell}) = \exp(-\jmath 2 \pi \eta_n \sin \vartheta _\ell/\lambda )$, $\eta_n$ is the coordinate of the $n$-th array element, and $\vartheta _{\ell}$ is the azimuth angle ranging from  $0^{\circ}$ to $180 ^{\circ}$. Therefore, the parameters to be estimated for each MPC under the ULA configuration are further denoted as $\Theta_\ell=\{\alpha_\ell,\vartheta _\ell,\tau_\ell\}$, $\ell=1,\dots,L$.

The NOMP algorithm is an iterative algorithm to extract the frequencies and amplitudes of a noisy mixture of sinusoids originally \cite{NOMP}. Extended versions of the NOMP estimator have been proposed to extract the angular and delay parameters of each MPC, and specific descriptions can be found in \cite{NOMPHan,NOMPTian}. Note that the number of MPCs $L$ to be estimated in the NOMP estimator should be predefined large enough to capture all significant paths. In this paper, we set $L=100$ to achieve a good tradeoff between accuracy and computational complexity \cite{HuangJSAC}, and the angular resolution in the NOMP algorithm is $1^{\circ}$. Therefore, the estimated spatial-temporal MPC parameters are used to further investigate important propagation channel characteristics. Meanwhile, the reconstructed channel matrix $\bH\in \bbC^{N_{\rm f}\times N}$ is also derived by synthesizing all estimated MPCs, i.e., 
\begin{equation}
    \left[ {\bH} \right] _{n_{\rm f},n} = \sum_{\ell=1} ^{L} \alpha_{\ell} e^{-\jmath 2\pi \eta_n \sin \vartheta _{\ell}/\lambda} e^{-\jmath 2\pi \eta _{\rm f} \Delta _{\rm f} \tau _{\ell}},
    \label{eq:channelrecon}
\end{equation}
for $n_{\rm f} = 1,\dots,N_{\rm f}$, $n=1,\dots,N$. 

\section{Analysis, Modeling, and Comparisons of U6G Channel Characteristics}
\label{sec:measure_result}

Based on the measurement setups described in previous section, we present the results and analyze the small-scale fading characteristics of the U6G frequency band, while also modeling parameters and comparing different scenarios. The analysis and discussions are based on measurement results derived from an array with dimension $N=8$. Note that in this section, we focus on the basic U6G channel characteristics under various scenarios. The channel characteristics presented here can also be regarded as those observed from a subarray. Furthermore, the channel characteristics of U6G ELAA and the relationships among different subarrays of an ELAA will be detailed in the next section.
  
\begin{figure*}[!t]
	\centering
    \subfloat[]{\includegraphics[width=.32\linewidth]{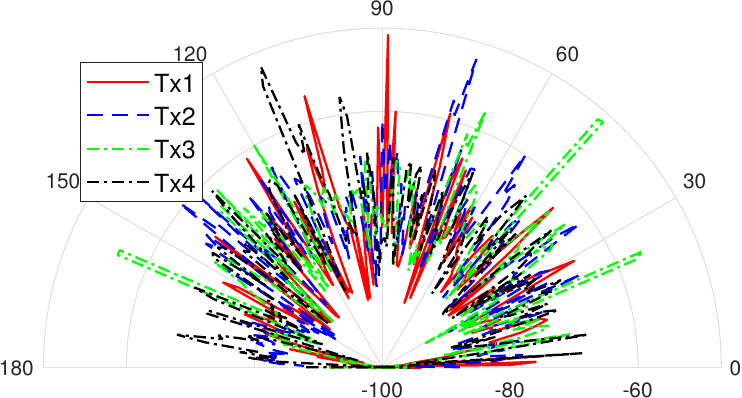}\label{fig:APAS_indoor}}\hspace{2pt}
	\subfloat[]{\includegraphics[width=.32\linewidth]{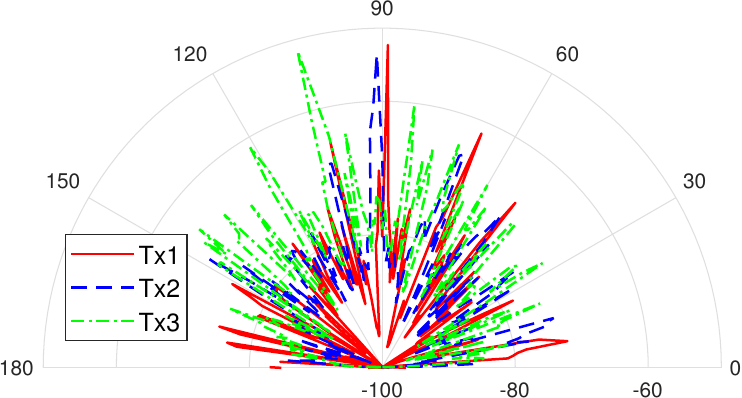}\label{fig:APAS_outdoor}}
    \subfloat[]{\includegraphics[width=.32\linewidth]{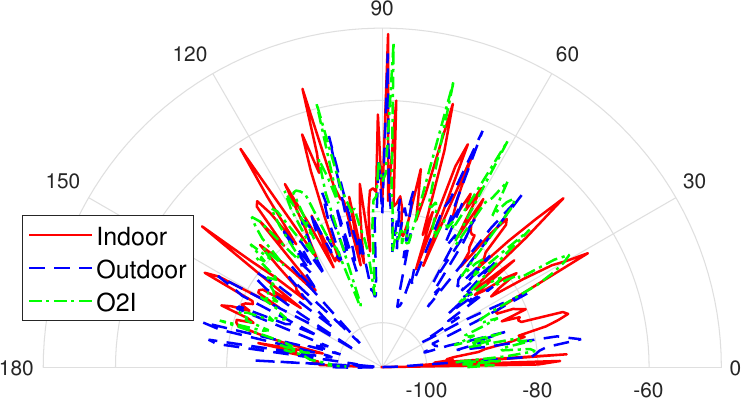}\label{fig:APAS_cmp}}
	\caption{APAS: (a) Indoor. (b) Outdoor. (c) Comparison of indoor (Tx1), outdoor (Tx1) and O2I.}
	\label{fig:APAS}
\end{figure*}

\subsection{APAS}
 
Based on the estimated MPC parameters, the APAS is obtained by
\begin{equation} 
    \text{APAS}(\vartheta) = \frac{1}{Q} \sum_{\ell=1}^{L}  \sum _{q=1} ^{Q}
 \big|\alpha_{\ell}^{(q)}\big|^2  \ \delta{\left(\vartheta-\vartheta_{\ell}^{(q)}\right)},
\end{equation}
where $Q$ is the number of PAS observations, $\alpha_{\ell}^{(q)}$ and $\vartheta _{\ell}^{(q)}$ are the complex amplitude and azimuth angle of the $\ell$-th MPC in the $q$-th observation, respectively.
The APAS reflects the power distribution among different angular directions \cite{PAS,covint}, which can be modeled using various distributions such as the Gaussian, Laplace, and Von-Mises distributions (VMD) \cite{PAS}, i.e.,
\begin{equation}
\text{APAS}(\vartheta)\!=\!
\begin{cases}
a_{\rmG}\!\times\! \exp {\left(-\left( \frac{\vartheta - b_{\rmG}} {c _{\rmG}} \right)^2\right)},&{\text{Gaussian}},\\
\frac{c _{\rmL}}{a _{\rmL} \sqrt{2\pi}} \!\times\! \exp {\left( - \frac{\sqrt{2}|\vartheta - b_{\rmL}| } {a _{\rmL}} \right)}, &{\text{Laplace}},\\
\frac{1}{2\pi I_0(c _{\rmV})}\!\times\! \exp {\left(c _{\rmV}\cos ( \vartheta - b _{\rmV})\right)}, & \text{VMD},
\end{cases}
\label{eq:APAS_distribution}
\end{equation}
where $\{a_{\rmX_1},b_{\rmX_2},c_{\rmX_2}\}$, $\rmX_1 \in \{\rmG,\rmL\}$, $\rmX_2 \in \{\rmG,\rmL,\rmV\}$ are the parameters of the above distributions. Note that $b_{\rmX}$, $\rmX \in \{\rmG,\rmL,\rmV\}$ represents the mean, and $c_{\rmX} \in \{\rmG,\rmL\}$ controls the width of the angular distribution.

In Fig. \ref{fig:APAS}, we compare the APASs under indoor, outdoor, and O2I scenarios.
It is evident that for each Tx location, the LoS direction is directed towards the Rx array location, while other MPCs correspond to the scattering environment.
From Fig. \ref{fig:APAS}\subref{fig:APAS_indoor}, we observe that for Tx positions 1 and 2, a single LoS path is dominant, while more MPCs appear for Tx positions 3 and 4.
This can be explained by the proximity of the Tx to the wall, which may generate more MPCs due to a more complex propagation environment.
In Fig. \ref{fig:APAS}\subref{fig:APAS_outdoor}, similar MPC characteristics are exhibited at Tx positions 1 and 2 in outdoor scenarios, reflecting a similar propagation environment.
Meanwhile, more MPCs are distinguishable at Tx location 3 due to slight blockages by trees.
Furthermore, when the Tx is facing the RX, i.e., Tx1 positions are considered for both indoor and outdoor scenarios, Fig. \ref{fig:APAS}\subref{fig:APAS_cmp} shows that similar MPC characteristics are observed under indoor and O2I scenarios, while fewer MPCs are seen under the outdoor scenario due to the open propagation environment.

Following the modeling ideology in \eqref{eq:APAS_distribution}, the measured data are fitted using both Gaussian and Laplace distributions, with parameters listed in Table \ref{tab:PAS}.
Note that we only capture the dominant components, i.e., MPCs corresponding to the LoS path, using a single distribution because the LoS link exists in our measurement scenarios.
Following the parameter in Table \ref{tab:PAS}, the MPC characteristics shown in the last paragraph can also be verified.
Moreover, the PAS can be modeled and configured for performance evaluations and simulations.

\begin{table*}[!t]\normalsize
    \centering
    \caption{Gaussian distribution and Laplace distribution parameters of PAS.}
    \label{tab:PAS}
    \begin{tabular}{|c|c|c|c|c|c|c|c|c|c|}
        \hline 
        \multicolumn{2}{|c|}{\multirow{2}{*}{Scenarios}} & \multicolumn{4}{c|}{Indoor} &\multicolumn{3}{c|}{Outdoor} & O2I \\
        \cline{3-10}
        \multicolumn{2}{|c|}{~}& Tx1 & Tx2 & Tx3 & Tx4 & Tx1 & Tx2 & Tx3 & Tx\\   
        \hline
        \multirow{3}{*}{Gaussian distribution} & $a_{\rmG}$ & 0.59  & 0.50  & 0.31  & 0.45   & 0.77  & 0.84  & 0.47   & 0.58 \\
                                             ~ & $b_{\rmG}$ & 89.00 & 73.24 & 48.46 & 112.30 & 88.95 & 91.16 & 104.90 & 88.03\\
                                             ~ & $c_{\rmG}$ & 0.39  & 0.49  & 1.46  & 0.74   & 0.41  & 0.61  & 0.69   & 0.42 \\
        \hline
        \multirow{3}{*}{Laplace distribution} & $a_{\rmL}$ & 1.16  & 0.38  & 0.70  & 0.63   & 0.25  & 0.72  & 0.59   & 0.23 \\
                                            ~ & $b_{\rmL}$ & 89.00 & 73.21 & 48.24 & 112.00 & 89.00 & 91.00 & 104.90 & 88.06\\
                                            ~ & $c_{\rmL}$ & 1.29  & 0.83  & 0.6   & 0.60   & 0.47  & 0.95  & 0.95   & 0.48 \\
        \hline 
    \end{tabular}
\end{table*}

\subsection{RMS AS}
Beyond the intuitive presentation above, the RMS AS under different scenarios is also compared. The RMS AS, a second-order statistic, describes the dispersion of the power angular spectrum and is calculated as follows: 
\begin{equation}   
	\vartheta_{\mathsf{rms}}=\sqrt{\frac{\sum_{\ell=1} ^{L} P _{\ell} \vartheta _{\ell} ^2}{\sum _{\ell=1} ^{L} P _{\ell}} - \left( \frac{ \sum _{\ell=1} ^{L} P _{\ell} \vartheta_{\ell} }{\sum_{\ell=1} ^{L} P _{\ell} } \right) ^2},
\end{equation}
where $P _{\ell} = |\alpha _{\ell}| ^2$ represents the power of the $\ell$-th MPC.

\begin{figure}[!t]
    \centering
    \includegraphics[scale=0.6]{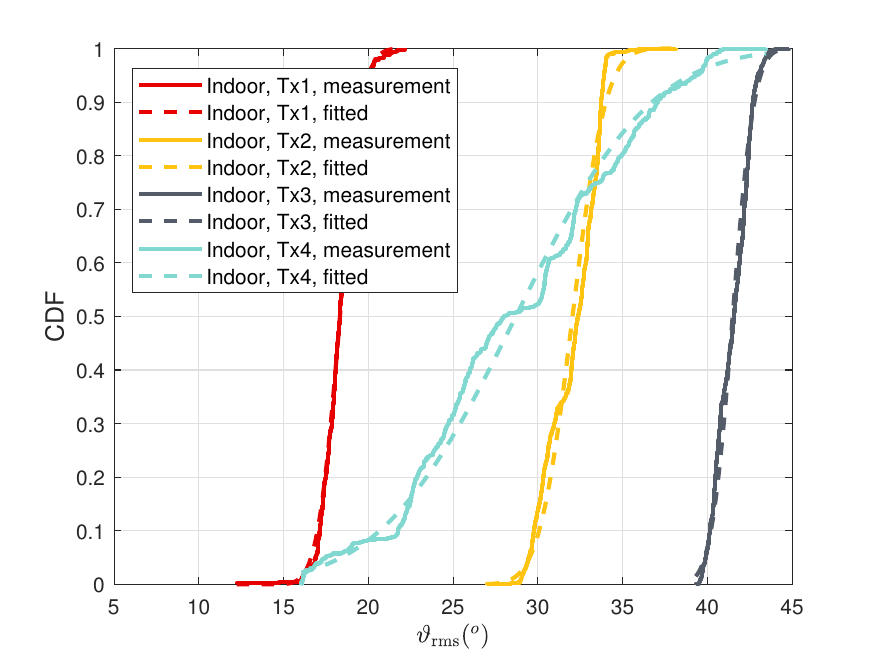}
    \caption{CDF of RMS-AS in indoor scenarios.}
    \label{fig:indoor_RMS_AS}
\end{figure}

\begin{figure}[!t]
    \centering
    \includegraphics[scale=0.6]{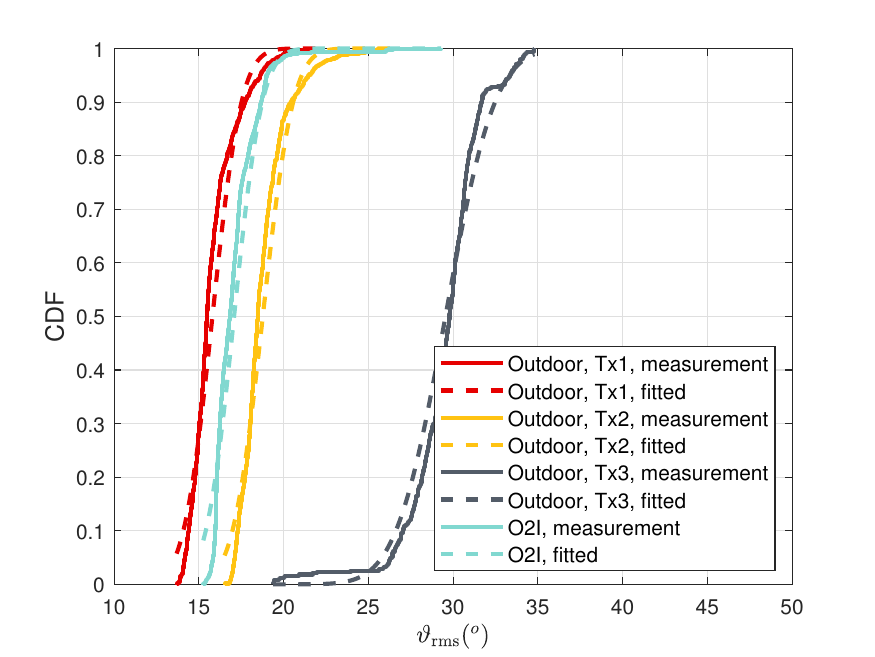}
    \caption{CDF of RMS-AS in outdoor and O2I scenarios.}
    \label{fig:outdoor_RMS_AS}
\end{figure}

Fig. \ref{fig:indoor_RMS_AS} displays the cumulative distribution functions (CDFs) of RMS AS for the four Tx locations in the indoor scenario. All measured data are fitted by a Gaussian distribution with parameters $\calN(18.31,1.13^2)$, $\calN(32.04,1.54^2)$, $\calN(41.49,1.01^2)$, and $\calN(28.70,6.27^2)$ for Tx1, Tx2, Tx3, and Tx4 respectively, in degrees. The good fit of these curves to the data confirms the appropriateness of the Gaussian model for RMS AS. It is evident that the AS values at Tx2, Tx3, and Tx4 are generally larger than those at Tx1, suggesting that a larger AS occurs when the Tx is not directly facing the Rx. The increased AS at Tx3 and Tx4 is due to the presence of more MPCs caused by walls or glass. 
Correspondingly, Fig. \ref{fig:scenarios}\subref{fig:indoorhall} suggests that the significant fluctuation of RMS AS values at Tx4 results from partial wall blockage. 

In Fig. \ref{fig:outdoor_RMS_AS}, the CDF of RMS AS and corresponding fitted curves for outdoor and O2I scenarios are compared. The parameters for the Gaussian distribution are $\calN(15.80,1.34^2)$, $\calN(18.77,1.40^2)$, $\calN(29.67,2.00^2)$, and $\calN(17.08,1.30^2)$ for outdoor Tx1, Tx2, Tx3, and O2I, respectively, in degrees. Compared to the red curves in Fig. \ref{fig:indoor_RMS_AS}, the AS under outdoor scenarios is generally similar to that under indoor scenarios when a LoS link exists between Tx and Rx. Comparing AS under Tx1 and Tx2, larger distances between Tx and Rx in complex scattering environments result in larger AS. The dark curves in Fig. \ref{fig:outdoor_RMS_AS} show heavy AS fluctuations caused by tree blockages. However, the AS under the O2I scenario is smaller due to the concentrated MPCs within a narrow angular range, constrained by the narrow door frame. 

\begin{figure*}[!t]
	\centering
    \subfloat[]{\includegraphics[width=.33\linewidth]{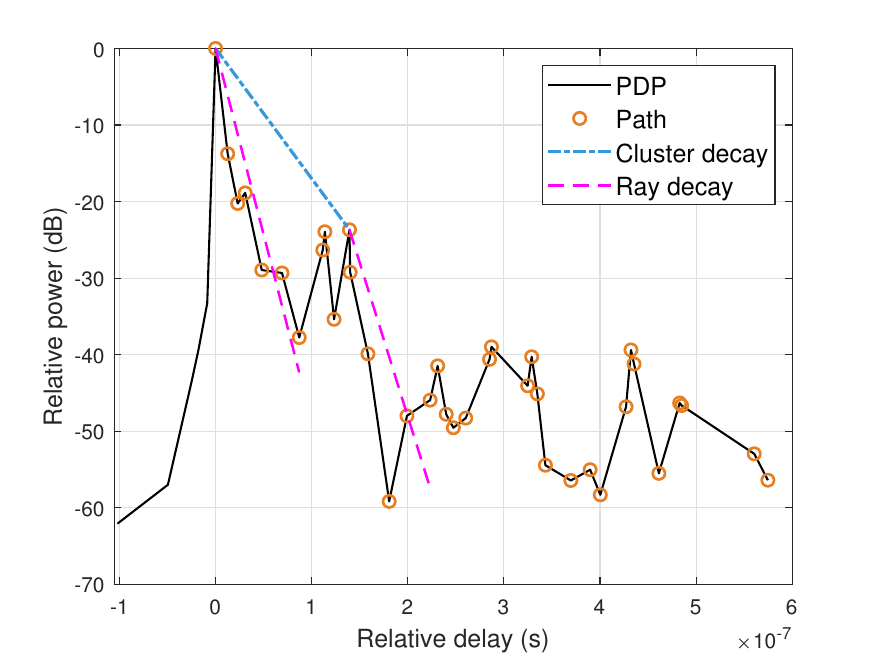}\label{fig:indoor_PDP}}
	\subfloat[]{\includegraphics[width=.33\linewidth]{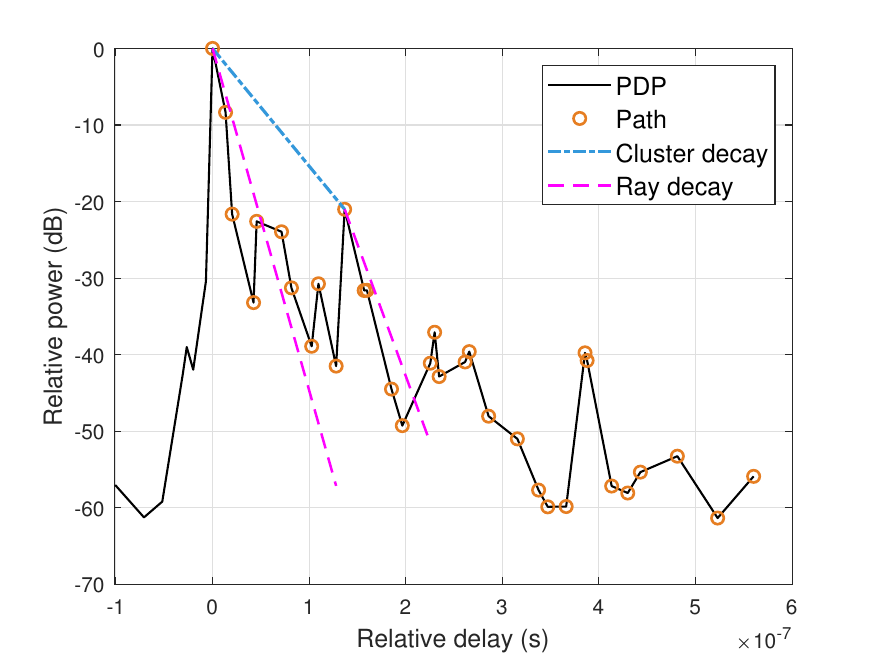}\label{fig:indoor_PDP2}}
    \subfloat[]{\includegraphics[width=.33\linewidth]{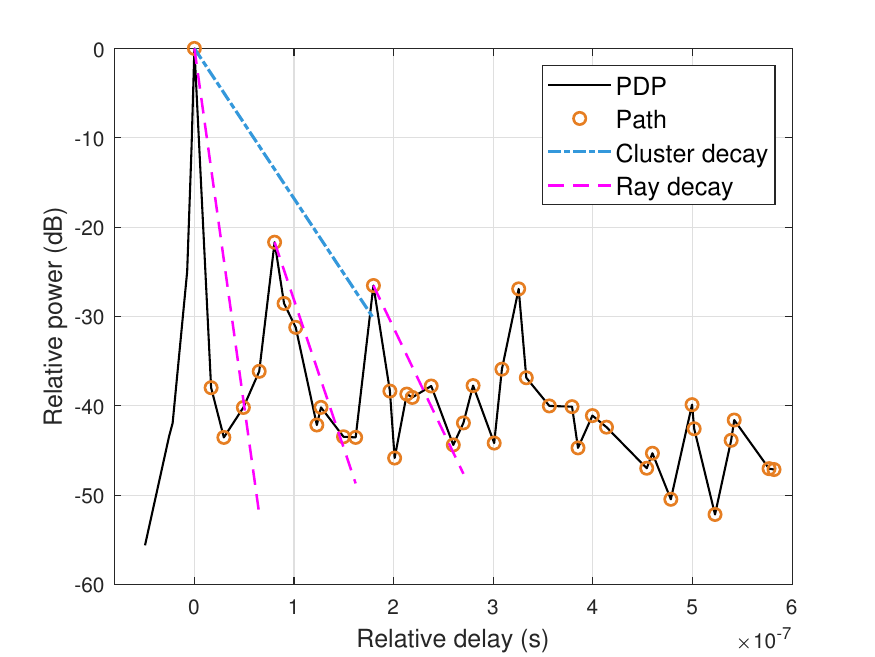}\label{fig:outdoor_PDP}}\\
    \subfloat[]{\includegraphics[width=.33\linewidth]{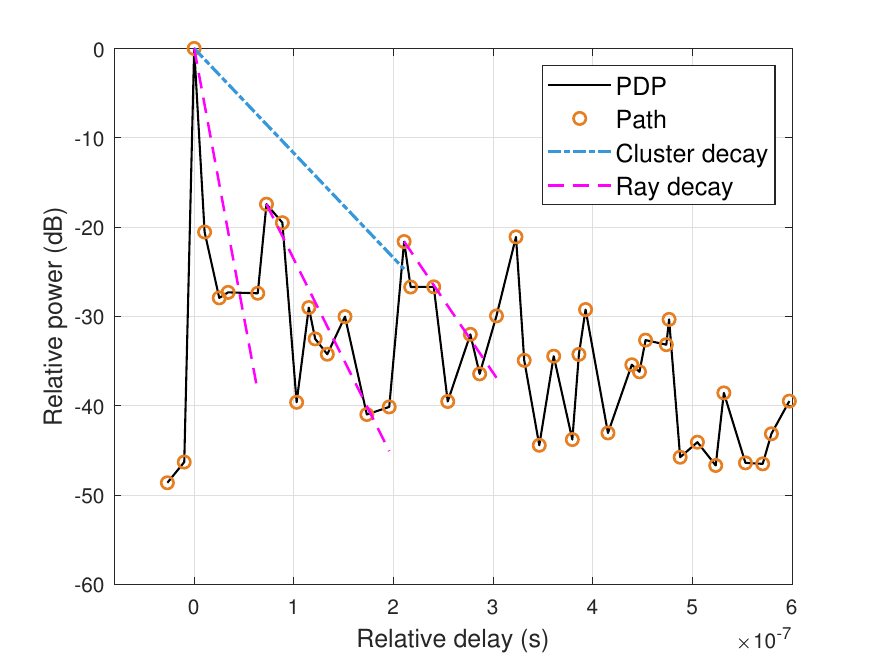}\label{fig:outdoor_PDP2}}
    \subfloat[]{\includegraphics[width=.33\linewidth]{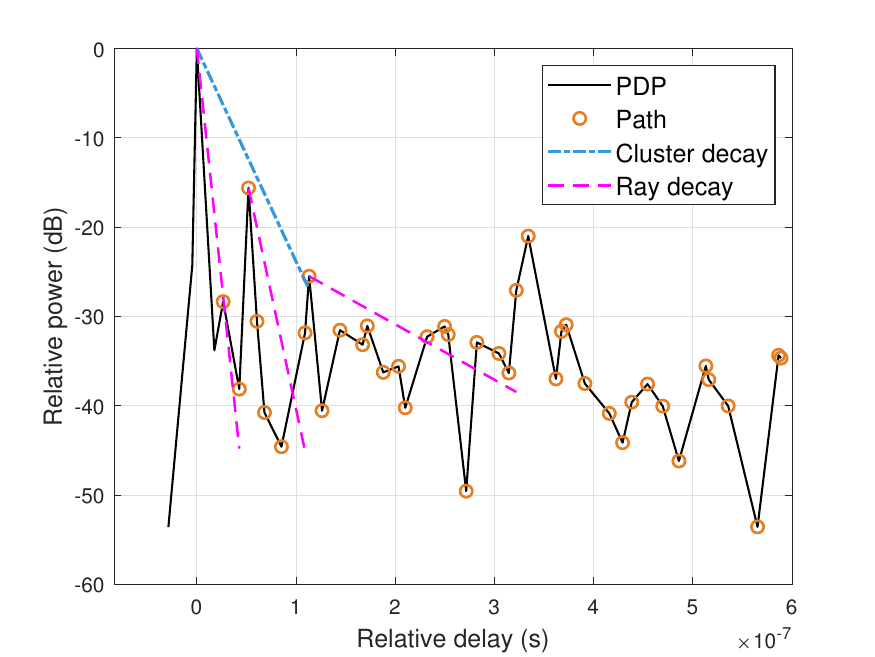}\label{fig:outdoor_PDP3}}
    \subfloat[]{\includegraphics[width=.33\linewidth]{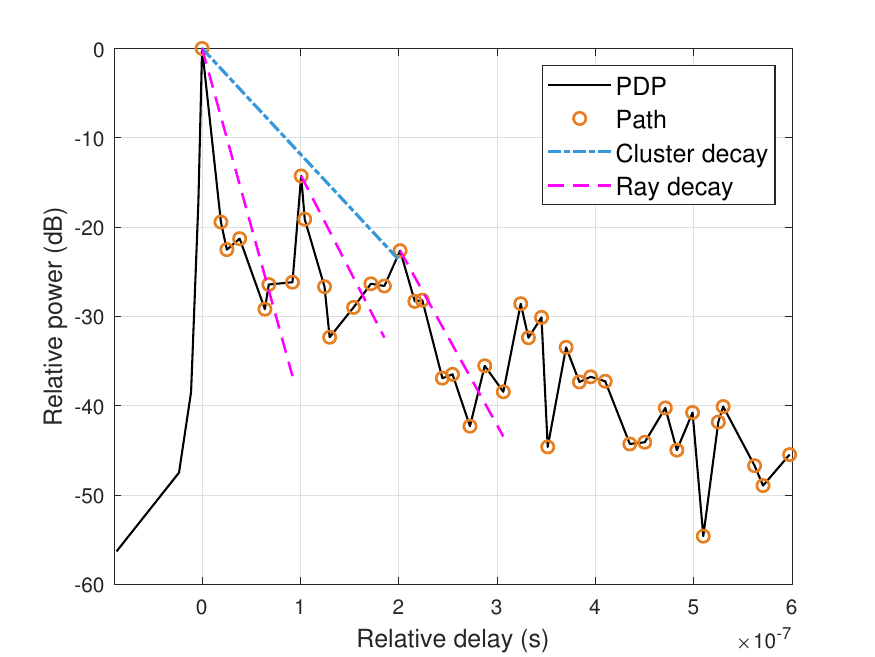}\label{fig:O2I_PDP}}
	\caption{PDP in Various Scenarios: (a) Indoor at Tx1, (b) Indoor at Tx2, (c) Outdoor at Tx1, (d) Outdoor at Tx2, (e) Outdoor at Tx3, (f) O2I.}
	\label{fig:PDP}
\end{figure*}

\begin{table*}[!t]\small
    \centering
    \caption{Cluster Parameters of different scenarios.}
    \label{tab:PDP}
    \begin{tabular}{|c|c|c|c|c|c|c|c|c|c|}
        \hline
        \multicolumn{2}{|c|}{\multirow{2}{*}{Scenarios}} & \multicolumn{4}{c|}{Indoor} &\multicolumn{3}{c|}{Outdoor} & O2I \\
        \cline{3-10}
        \multicolumn{2}{|c|}{~}             & Tx1 & Tx2 & Tx3 & Tx4 & Tx1 & Tx2 & Tx3 & Tx\\ 
        \hline
        \multicolumn{2}{|c|}{Cluster decay factor [ns]} & 25.56  & 28.26  & 18.04  & 18.22  & 25.84  & 37.05  & 18.21   & 36.71 \\
        \hline
        \multirow{3}{*}{Ray decay factor [ns]}  & Cluster 1 & 8.94  & 9.72  & 7.75  & 8.16  & 5.42  & 7.21  & 4.15   & 10.88 \\
                                  ~ & Cluster 2 & 10.79 & 12.62 & 31.83 & 19.46 & 13.08 & 19.34 & 8.41 & 20.31\\
                                  ~ & Cluster 3 & -- & -- & --  & -- & 18.62 & 26.39  & 69.96   & 21.91 \\
        \hline
        \multicolumn{2}{|c|}{Power ratio of cluster 1} & 99.35\%  & 99.04\%  & 99.70\%  & 99.16\% & 91.16\%  & 94.51\%  & 94.90\% & 91.87\% \\
        \hline
    \end{tabular}
\end{table*}

\subsection{PDP}
Next, we focus on the small-scale fading in the delay domain. The PDP, derived from parameters extracted using the NOMP algorithm, is expressed as follows:
\begin{equation}
    \text{PDP}(\tau) = \sum _{\ell =1} ^{L} P _{\ell} \delta (\tau - \tau _\ell),
\end{equation}
The PDP for indoor, outdoor, and O2I scenarios is illustrated in Fig. \ref{fig:PDP}, where the absolute power and delay are normalized by the maximum and corresponding delay, respectively. This normalization helps focus on the decay characteristics with respect to the delay. 

Observing the PDP, we analyze the U6G channel from a cluster perspective, i.e., the Saleh-Valenzuela (S-V) model \cite{S-V}, as follows:
\begin{equation}  
    h(\tau) = \sum _{m = 1} ^{M} \sum _{r_m=1} ^{R_m} \alpha _{m,r_m}e ^{j\varphi _{m,r_m}} \delta(\tau - \Gamma _m - \tau _{m,r_m}),
\end{equation}
where $M$ and $R_m$ represent the number of clusters and rays in the $m$-th cluster, respectively, and $\sum _{m=1} ^{M} R_m = L$; $\Gamma _m$ is the arrival delay of the $m$-th cluster; $\tau _{m,r_m}$ is the relative delay of the $r_m$-th ray in the $m$-th cluster; and $\varphi _{m,r_m}$ is the phase of the $r_m$-th ray. Considering the decay characteristics, the power of the $r_m$-th ray is modeled as 
\begin{equation}  
{P_{m,r_m}}={P_{0,0}} \exp{\left(-\frac{\Gamma_m}{\bar{\Gamma}}\right)}  \exp{\left(-\frac{\tau_{m,r_m}}{\bar{\tau} _m}\right)},
\label{eq:fading}
\end{equation}
where ${P_{m,r_m}}=|\alpha _{m,r_m}|^2$ denotes the average power of the $r_m$-th ray in the $m$-th cluster; $\bar{\Gamma}$ and $\bar{\tau} _m$ represent the decay factors for power within clusters and for rays in cluster $m$, respectively; and ${P_{0,0}}$ is the initial average power of the first ray in the first cluster. A comprehensive illustration of these decay factors is provided in Table \ref{tab:PDP}. 

In the indoor scenarios, two clusters capture the dominant MPCs, while three clusters are typical in the outdoor and O2I scenarios. In indoor environments, channel power is predominantly concentrated in the first cluster due to the relatively simple structure of indoor halls. For outdoor scenarios, the ray decay factors tend to increase with the number of clusters, reflecting a slower decay trend caused by multiple rays with comparable power within the second and third clusters. As observed in Fig. \ref{fig:PDP}\subref{fig:O2I_PDP}, in the O2I scenario, the decay of each cluster follows an exponential formula, and the ray decay factors of the latter clusters are similar. This alignment with the S-V model illustrates that the measured PDP in the O2I scenario reflects a typical scattering environment caused by multiple clusters.    

\subsection{RMS DS}
  
\begin{figure}[!t]
    \centering
    \includegraphics[scale=0.6]{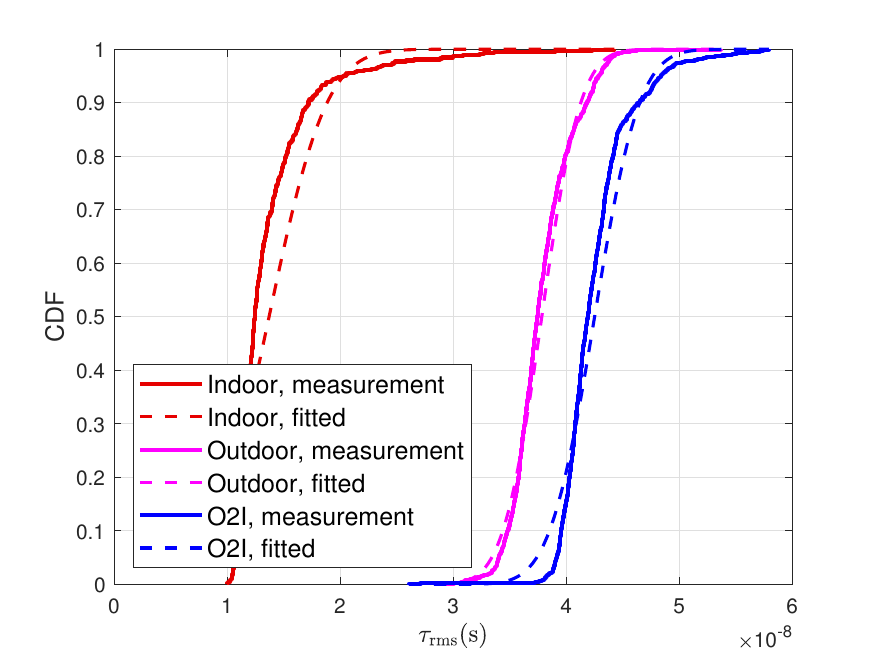}
    \caption{CDF of RMS DS in indoor, outdoor, and O2I scenarios.}
    \label{fig:RMS_DS}
\end{figure}

The RMS DS is a second-order statistic used to describe the dispersion of the power delay profile, calculated as
\begin{equation}
    \tau _{\mathsf{rms}} = \sqrt{\frac{\sum_{\ell=1} ^{L} P _{\ell} \tau _{\ell} ^2}{\sum _{\ell=1} ^{L} P _{\ell}} - \left( \frac{ \sum _{\ell=1} ^{L} P _{\ell} \tau _{\ell} }{\sum_{\ell=1} ^{L} P _{\ell} } \right) ^2}.
\end{equation}
In Fig. \ref{fig:RMS_DS}, the CDFs of RMS DS and corresponding fitted curves for indoor, outdoor, and O2I scenarios are compared. Tx1 positions were selected for both indoor and outdoor scenarios to ensure direct transmission facing in different environments. Note that all fitted curves align well with the measured data, validating the appropriateness of the Gaussian model for RMS DS. 
The parameters of the Gaussian distribution are $\calN(14.36,4.96^2)$, $\calN(37.64,2.80^2)$ and $\calN(42.52,3.12^2)$ for indoor, outdoor, and O2I scenarios, respectively, with units in nanoseconds (ns). In indoor scenarios, DS values are concentrated within the range of 10 to 20 ns, consistent with results in \cite{RBCM2}. Meanwhile, DS values span wider ranges of 30 to 45 ns and 35 to 50 ns for outdoor and O2I scenarios, respectively, due to longer propagation distances and more complex environments.

\subsection{Rician K Factor}

\begin{figure}[!t]
    \centering
    \includegraphics[scale=0.6]{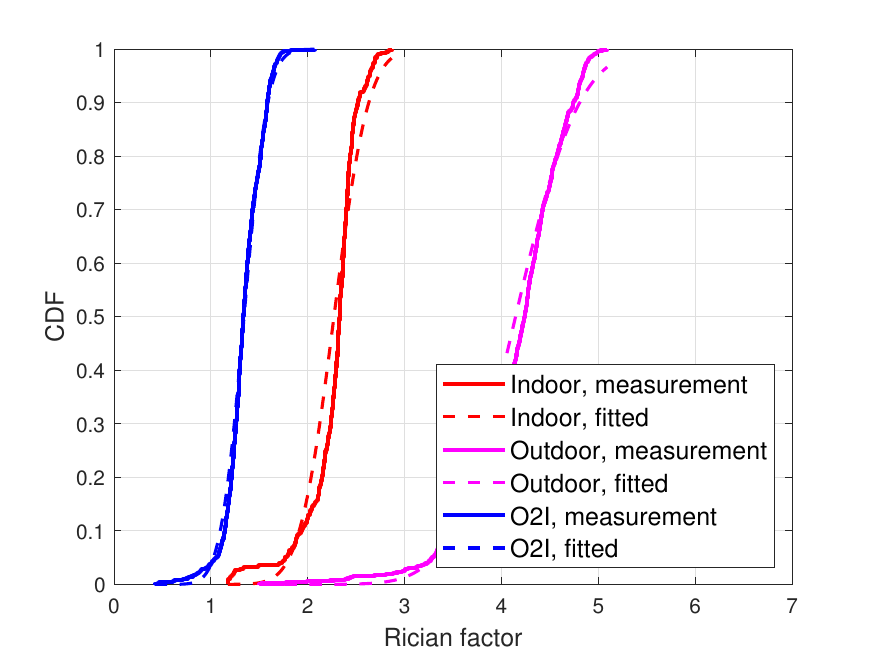}
    \caption{CDF of the Rician K factors in indoor, outdoor, and O2I scenarios.}
    \label{fig:RicianK}
\end{figure}
 
The Rician K factor is a crucial indicator of narrowband fading characteristics. We calculate the Rician factor based on the definition provided in \cite{YinRician}. In the LoS environment, the power of the LoS path is maximal among all estimated MPCs, while the power of the NLoS components is aggregated from the remaining MPCs. The powers of the LoS is denoted as $P_{\LoS}$. The Rician K factor is then calculated by 
\begin{equation}
    K = \frac{P_{\LoS}}{P_{\mathrm{tot}} - P_{\LoS}},
\end{equation}
where $P _{\mathrm{tot}}$ is the total power of all estimated MPCs.

Fig. \ref{fig:RicianK} shows the CDF of Rician K factors under different scenarios, with Tx1 positions selected for both indoor and outdoor scenarios to ensure directly facing transmissions across different environments. The measured Rician factors are fitted using Gaussian distributions, with fitted parameters  $\calN(2.30,0.28^2)$, $\calN(4.14,0.52^2)$, and $\calN(1.34,0.19^2)$ for the indoor, outdoor, and O2I scenarios, respectively. The fitted curves align closely with the measured data, further validating the Gaussian modeling of the Rician factor. It is observed that the Rician factors under the outdoor scenario are higher than those under the indoor and O2I scenarios when all are under strong LoS conditions. The dominance of the LoS path in the outdoor scenario, as reflected by the Rician K factor, is consistent with earlier findings. However, due to more MPCs caused by the complex scattering environment, the Rician K factors experience degradation in the O2I scenario.

\subsection{Sparsity}

\begin{figure}[!t]
\centering
\includegraphics[scale=0.6]{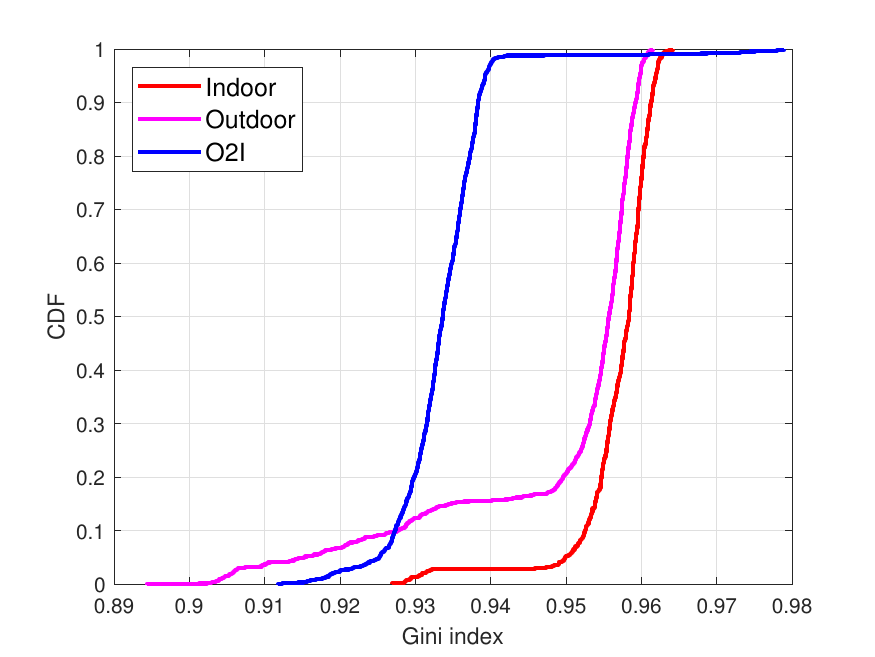} 
\caption{CDFs of the Gini index at Tx1 positions in various scenarios.}
\label{fig:Gini_index}
\end{figure}

To evaluate the sparsity of the U6G channel, we adopt the Gini index as a measure \cite{sparsity1,sparsity2}. Specifically, we consider a scenario where $L$ MPCs exist, denoted threr channel impulse power vector as $ [P_1,\dots,P_L]$. These powers, $P_\ell$, where $\ell = 1,\dots,L$, are ordered from the smallest to the largest, i.e., $P _{(1)}\leq \dots \leq P _{(L)}$, where $(\ell)$  indicates the new index after a sorting operation. The Gini index is then calculated by:
\begin{equation}  
    G = 1 - 2 \sum _{\ell = 1} ^{L} {\left( \frac{P_{(\ell)}}{P _{\mathrm{tot}}  } \cdot \frac{L-\ell + 0.5}{L}\right)}.
\end{equation}
The Gini index, meeting the definition of sparsity, thus serves as an effective indicator of this property. Note that the Gini index ranges from 0 to 1, where $G = 0$ represents a scenario in which all MPCs possess equal power, and $G = 1$ indicates the sparsest case, where all the power is concentrated on a single MPC.

Fig. \ref{fig:Gini_index} compares the CDFs of the Gini index under position Tx1 across different scenarios. It can be observed that the Gini indices are predominantly concentrated in the range from 0.91 to 0.96, indicating that the U6G channel exhibits sparsity under different scenarios when a LoS link exists. As shown, channel sparsity declines in the O2I scenario due to increased reflection and scattering from the complex propagation environment. However, both indoor and outdoor scenarios demonstrate significant sparsity in the U6G channel when a LoS link is present.

\section{Distributed Channel Characteristics of U6G ELAA Systems} 
\label{sec:ELAA} 

Following the analysis and comparisons of basic channel characteristics, this section explores the channel characteristics of the entire ELAA. These results are derived from virtual array measurements conducted in an indoor scenario, where the Rx is fixed and is positioned as depicted in Fig. \ref{fig:scenarios}\subref{fig:hall}, while the Tx is located approximately 5.6 meters in front of the Rx.
By horizontally moving the physical array with dimension $N = 8$, a virtual ELAA with dimension $N=64$ is created under the static measurement environment.

The primary goal of our measurements is to provide supportive propagation characteristics that can help reduce complexity and hardware costs through subarray-wise distributed processing. Therefore, the characteristics of the U6G ELAA channel are presented and analyzed primarily from the perspectives of subarrays and sub-bands. Specifically, we focus on the non-stationarities and consistencies of channel characteristics among different subarrays and sub-bands within a U6G ELAA system. Additionally, the theoretical subarray-wise far-field approximation is also verified.
 
\begin{figure}
    \centering
    \includegraphics[scale=0.6]{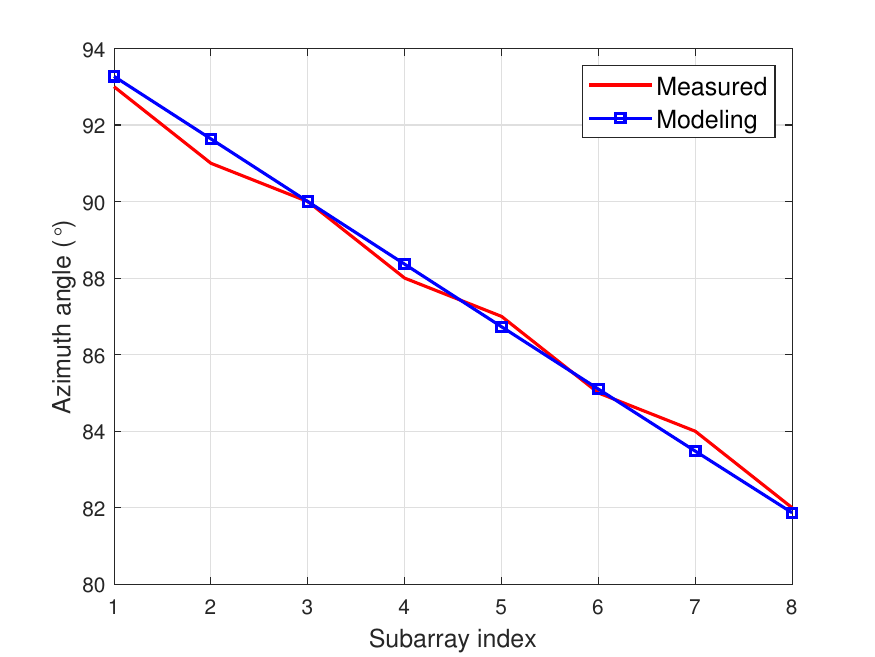}
    \caption{Azimuth angle variations versus subarrays.}
    \label{fig:AoA_subarray}
\end{figure} 

\subsection{Subarray-wise Non-stationarities}

Spatial non-stationarities refer to the phenomenon where different parts of the ELAA exhibit distinct propagation characteristics, both in large-scale and small-scale aspects, as confirmed by previous measurement campaigns \cite{HuangJSAC,NS,JHZhangNS}. In this study, we focus on non-stationarities in the angular domain, considering different uniformly divided subarrays. The analysis begins by comparing the azimuth angles of LoS paths among these subarrays.
For each subarray, the azimuth angle of the LoS path is selected from the extracted MPCs, which have the strongest power.
As depicted in Fig. \ref{fig:AoA_subarray},  the measured azimuth angles of the LoS paths for eight subarrays are presented, alongside theoretical modeling based on the spherical propagation model and geometric relationships such as those described in \cite{NFmodel}. The variation in azimuth angles along the ELAA, coupled with the close alignment between measured and theoretical azimuth angles, not only validates the near-field propagation characteristics of an ELAA but also demonstrates effective variability in azimuth angles across different subarrays.

\begin{table*}[!t]\normalsize
    \centering
    \caption{Gaussian distribution of RMS AS and Rician K factor Subarray-Wise.}
    \label{tab:RMSAS_subarray}
    \begin{tabular}{|c|c|c|c|c|c|c|c|c|}
        \hline
        Subarray index & 1 & 2 & 3 & 4 & 5 & 6& 7 & 8\\
        \hline
        Mean of RMS-AS (Unit: degree) & 11.00 & 14.43 & 11.78 & 13.80 & 19.27 & 17.69 & 17.16 & 20.73 \\
        Standard deviation of RMS AS (Unit: degree) & $1.96$ & $0.83$ & $0.29 $ & $1.03$ & $2.88$ & $1.75$ & $1.12$ & $0.49$\\
        \hline
        Mean of Rician K factor & 3.61 & 2.50 & 3.20 & 3.48 & 1.88 & 2.81 & 3.07 & 2.69 \\
        Standard deviation of Rician K factor & $0.27$ & $0.40$ & $0.27$ & $0.25$ & $0.84$ & $0.41$ & $0.32$ & $0.27$\\
        \hline
    \end{tabular}
\end{table*}

Next, we examine the RMS AS and the Rician K factor across various subarrays. Both the RMS AS values and Rician K factors for different subarrays have been fitted using Gaussian distributions, with parameters detailed in Table \ref{tab:RMSAS_subarray}. From Table \ref{tab:RMSAS_subarray}, it is evident that the RMS AS varies among the subarrays of an ELAA, with a noticeable trend that larger angles between the azimuth and the normal direction correlate with larger RMS AS values, aligning with findings from Fig. \ref{fig:indoor_RMS_AS}. Moreover, according to the means and variances of Rician K factors listed in Table \ref{tab:RMSAS_subarray}, the Rician K factors across different subarrays typically range from 1.5 to 4. Although the subarrays share a similar propagation environment, the dominance of the LoS path among MPCs slightly differs across the 8 subarrays.

\begin{figure}[!t]
    \centering
    \includegraphics[scale=0.6]{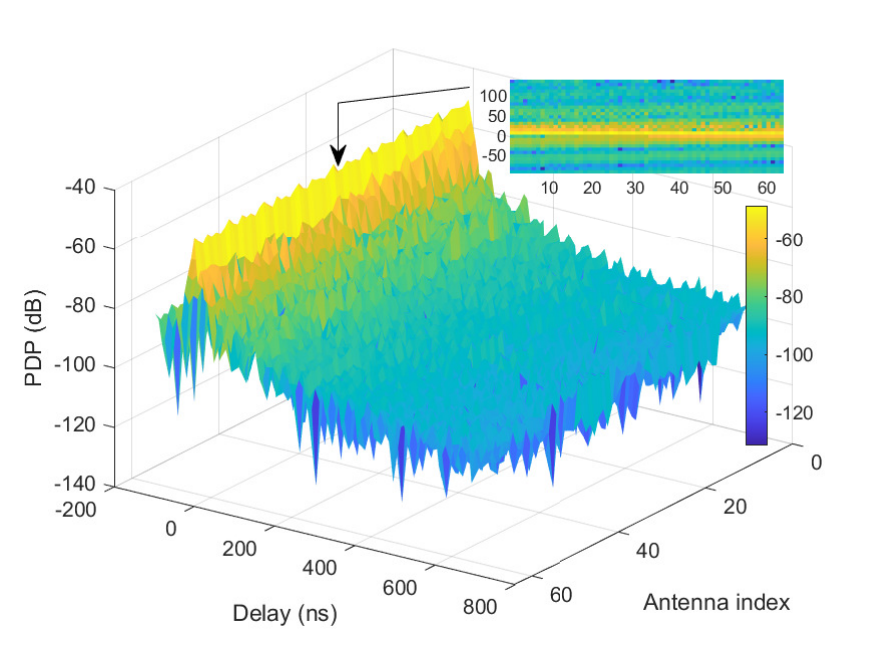}
    \caption{A snapshot of PDP of different subarrays.}
    \label{fig:PDP_virt}
\end{figure}

In terms of delay characteristics, we derive the PDP observed by 64 digital channels, as illustrated in Fig. \ref{fig:PDP_virt}. The power of the strongest multipath component, specifically the LoS path, exhibits variations across the ELAA elements, with a maximum difference of approximately 4.61 dB. This variation further substantiates the non-stationarities within the array.

\begin{figure}[!t]
	\centering
    \subfloat[]{\includegraphics[width=0.95\linewidth]{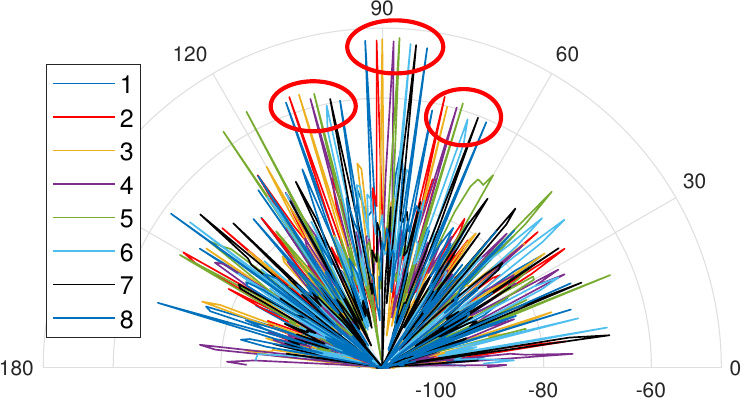}\label{fig:PAS_virt}}\\
	\subfloat[]{\includegraphics[width=1\linewidth]{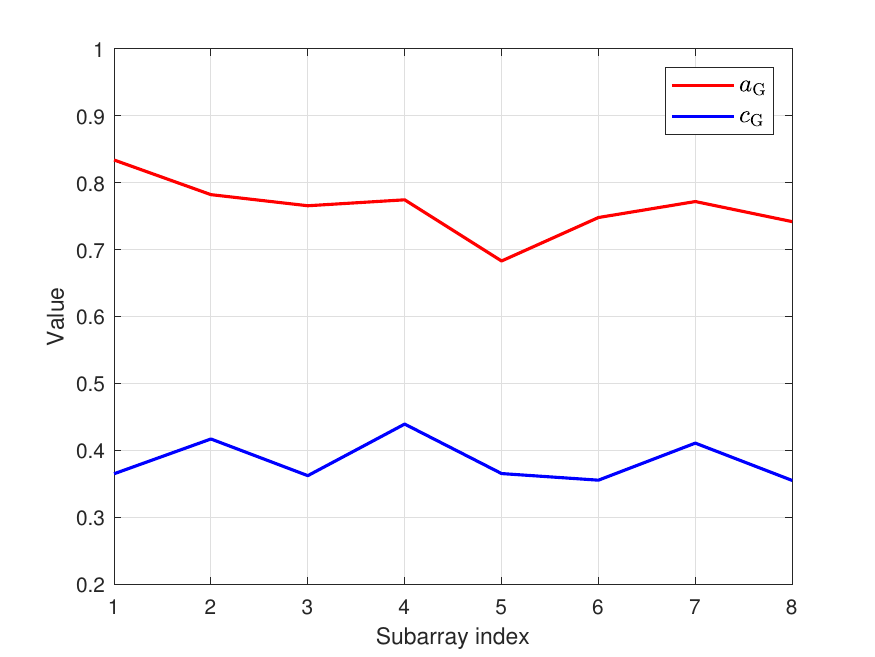}\label{fig:PAS_virt_fit}}
 \caption{Angular characteristics of different subarrays. (a) APAS of different subarrays. (b) Means and variances of the APAS values across each subarray.}
 \label{fig:PAS_sub}
 \end{figure}

\subsection{Subarray-wise Consistencies}
From the analysis above, we observed that different subarrays experience distinct propagation channel characteristics. However, to effectively implement distributed processing for ELAA systems, it is crucial to investigate the consistencies among subarrays. This is because all subarrays derive from the same ELAA channel, and therefore, some similarities should be present among them. Additionally, subarray-wise consistencies can support the deployment of identical hardware structures across subarrays in an ELAA, which is beneficial for the system's scalability.

To this end, we first analyze the APAS of different subarrays, as shown in Fig. \ref{fig:PAS_sub}\subref{fig:PAS_virt}. The dominant angles of different subarrays, marked by red circles, show slight variations but are consistent with the results in Fig. \ref{fig:AoA_subarray}. However, approximately three dominant angular directions are distinguishable across all eight subarrays, and these directions are similar to those in other subarrays. These findings confirm that similar MPC characteristics are shared among different subarrays. Subsequently, the APAS for different subarrays is fitted with a Gaussian distribution, and the amplitudes and variances are compared in Fig. \ref{fig:PAS_sub}\subref{fig:PAS_virt_fit}. We find that the means are roughly concentrated in the range of 0.7 to 0.8, while the variances are approximately within the range of 0.36 to 0.43. These results indicate that the distributions, including the amplitudes and spread of the LoS among MPCs, are consistent across different subarrays.
 
Regarding consistencies in the delay domain, we observe from Fig. \ref{fig:PDP_virt} that the delay of the LoS path is similar across the ELAA elements. This similarity is significant, especially when considering that the results are derived from a detection signal with a bandwidth of 100 MHz. It can be concluded that when an ELAA system operates with a moderate transmission bandwidth, the differences between different ELAA elements in the delay domain are not pronounced. However, when employing a wider bandwidth, the non-stationarities in the delay domain may become more distinguishable.

\begin{figure*}[!t]
	\centering
     \subfloat[]{\includegraphics[width=.64\linewidth]{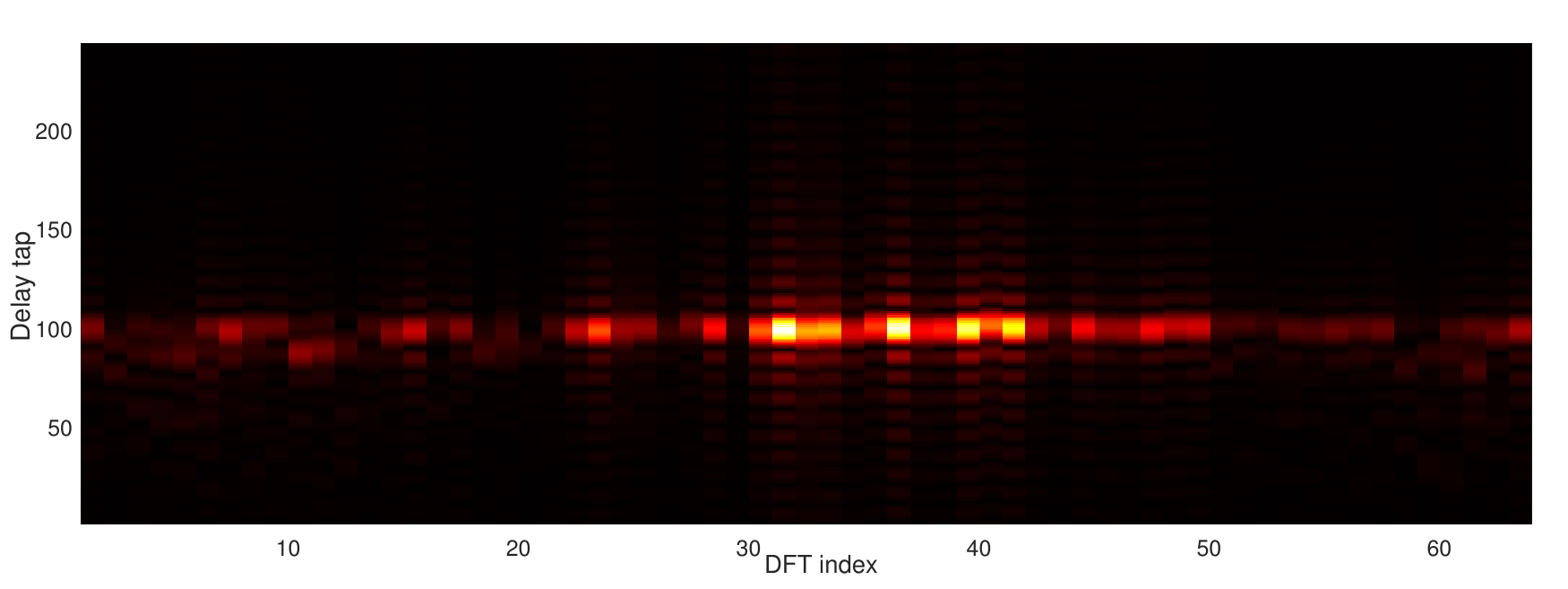}\label{fig:DFT_virt}}\hspace{2pt}
    \subfloat[]{\includegraphics[width=.32\linewidth]{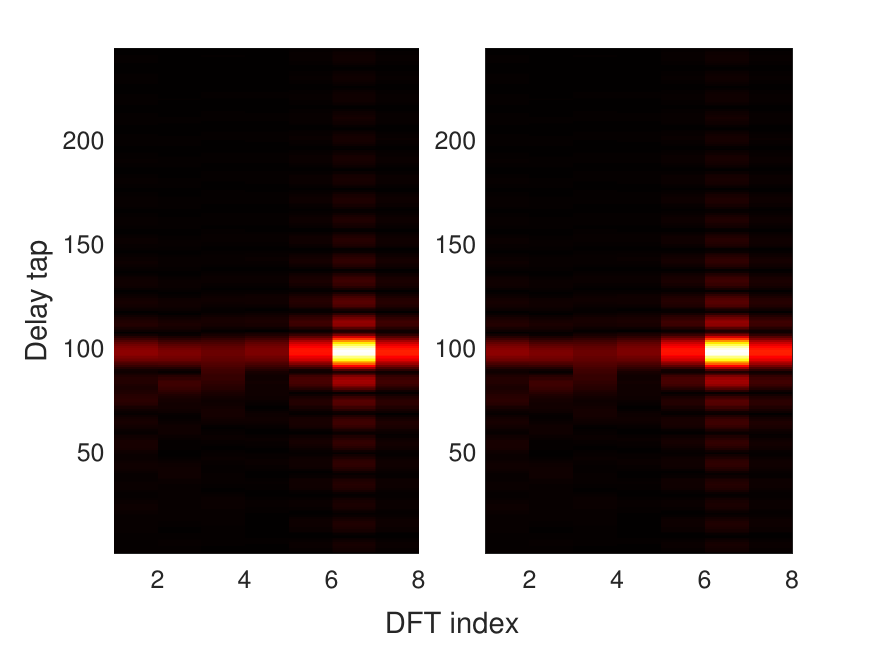}\label{fig:DFT_subarray}}
	\\
	\caption{Comparison of channel characteristics in delay and spatial domains after DFT: (a) The whole ELAA. (b) Subarray 1 and Subarray 6.}
	\label{fig:DFT_virt_subarray}
\end{figure*}

\subsection{Subarray-wise Far-field Approximation}
The previous subsections explored the similarities and differences across different subarrays. Next, we aim to understand the relationship between the whole ELAA and its subarrays under the framework of distributed signal processing. As discussed in Section \ref{sec:relatedworks}, an effective approach is to divide the ELAA into subarrays and treat each subarray's channel as a far-field approximation. This method helps manage the complex channel characteristics and high computational demands of ELAA systems. 

It is well-established that DFT vectors can serve as bases for far-field channels in both the delay and angular domains \cite{DFTLee,DFTGao}. Typically, the channel in the delay and spatial domain representation, denoted as $\bH _{\rmD\text{-}\rmS}$, is obtained by
\begin{equation}
    \bH _{\rmD\text{-}\rmS} = \bU_{\rmF} \bH \bU_{\rmA} ^{\ctrans},
\end{equation}
where $\bU _{\rmF} \in \bbC ^{N_{\rm f} \times N_{\rm f}}$ and $\bU _{\rmA} \in \bbC ^{N\times N}$ are the DFT matrices in the delay and angular domains, respectively.

We then compare the power of entries in $\bH_{\rmD\text{-}\rmS}$ under the whole ELAA and a subarray, i.e., $N=64$ and $N=8$, as drawn in Fig. \ref{fig:DFT_virt_subarray}\subref{fig:DFT_virt} and \ref{fig:DFT_virt_subarray}\subref{fig:DFT_subarray}, respectively.
Taking subarrays 1 and 6 as examples (Fig. \ref{fig:DFT_virt_subarray}\subref{fig:DFT_subarray}), it is observed that the power of a subarray channel is concentrated on almost one DFT direction in the spatial domain. Conversely, the power of the ELAA channel is distributed over several consecutive spatial DFT directions, with a more apparent spread among different DFT directions. These results not only validate the spread of near-field channels across multiple spatial angles but also demonstrate the effectiveness of the subarray-wise far-field approximation.

Based on \eqref{eq:channelrecon}, the subarray-wise far-field channel can be reconstructed. The error of the subarray-wise far-field approximation is then measured through the normalized mean square error (NMSE), defined as:
\begin{equation} 
    \text{NMSE} = \frac{1}{Q} \sum _{q=1} ^{Q} \frac{\| \bH_{\mathrm{virt}} ^{(q)} - \bH _{\mathrm{recon}} ^{(q)} \Vert ^2 _{\rm F}}{\| \bH_{\mathrm{virt}} ^{(q)} \Vert ^2 _{\rm F}},
\end{equation}
where $Q$ represents the number of observations, $\bH_{\mathrm{virt}} ^{(q)}$ is the channel matrix derived from the virtual array of the $q$-th observation, and $\bH _{\mathrm{recon}} ^{(q)} = [\bH _{\mathrm{recon},1} ^{(q)},\dots,\bH _{\mathrm{recon},B} ^{(q)}]$ is the reconstructed channel matrix under the subarray-wise far-field assumption for the $q$-th observation. Here, $B$ denotes the number of subarrays.
Specifically, the channel of subarray $b$, $\bH _{\mathrm{recon},b} ^{(q)}$, is derived based on the NOMP algorithm under the assumption of far-field conditions.
When the array is divided into $B=16$ and $B=8$ subarrays, each equipped with 4 and 8 array elements respectively, the NMSEs are -18.82 dB and -18.09 dB. These findings indicate that increasing the number of subarrays enhances the accuracy of the subarray-wise far-field approximation. Consequently, this supports the design of channel estimation and transmission strategies with reduced complexity.

\subsection{Sub-Band Characteristics}
Beyond the novel angular characteristics introduced by the increased array elements, U6G ELAA systems also feature a wide transmission bandwidth. Distributed processing techniques, such as out-of-band extrapolation of in-band channel characteristics \cite{NOMPHan,NOMPChen}, serve as effective means to reduce computational complexity. Therefore, this subsection investigates the wideband characteristics, particularly the angular characteristics, of different sub-bands within the U6G band.

\begin{figure}[!t]
    \centering
    \includegraphics[scale=0.6]{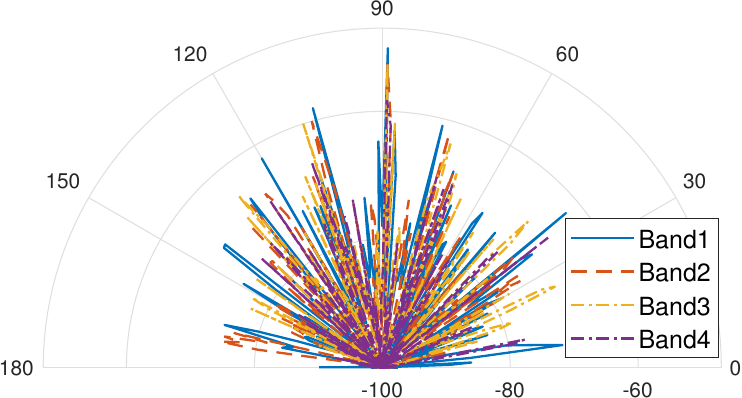}
    \caption{APAS of different sub-bands.}
    \label{fig:multiband_PAP}
\end{figure}

Taking the indoor Tx1 scenario as an example, we focus on four non-overlapping bands with bandwidths of $819 \Delta f$, specifically $0-818\Delta f$, $819-1637\Delta f$, $1638-2457\Delta f$, and $2458-3276 \Delta f$, denoted as Band1, Band2, Band3, and Band4, respectively. We begin by examining the APAP derived from these different sub-bands. As depicted in Fig. \ref{fig:multiband_PAP}, we compare the APAP across these sub-bands. It is evident that similar APAP patterns are observed under different sub-bands, with dominant angular directions appearing consistently across them. Specifically, the directions of the LoS paths under different sub-bands remain nearly identical, while other dominant directions exhibit slight deviations.

Next, we analyze the RMS AS derived from these sub-bands. The RMS AS values for Band1, Band2, Band3, and Band4 are $18.85^\circ$, $18.13^\circ$, $19.25^\circ$, and $18.73^\circ$, respectively. These results suggest that the RMS AS is relatively consistent across different sub-bands. However, a trend is observed where employing a wider band results in a larger RMS AS, potentially due to the increased number of MPCs observed and distinguished through high-dimensional joint frequency-spatial path extraction.

\subsection{Insights}
In this subsection, we summarize the channel characteristics of U6G ELAA systems based on the measurement results from Sections
\ref{sec:measure_result} and \ref{sec:ELAA}. The key insights can be concluded as follows:

\subsubsection{Angular Characteristics} 
    The power of the MPCs tends to concentrate in several distinct angular directions. When a LoS link exists between the Tx and the Rx, the LoS path is significantly more dominant than other MPCs. Comparing different propagation scenarios, additional MPCs tend to emerge in the angular domain under indoor and O2I scenarios, indicating richer scattering environments.

\subsubsection{Delay Characteristics}
    Cluster scattering properties are observable, where the power of MPCs tends to concentrate in a single cluster in indoor scenarios, while multiple clusters emerge in outdoor and O2I scenarios. The measured results align well with the S-V model, further validating the cluster scattering characteristics under the U6G frequency band.
 
\subsubsection{Subarray Characteristics}
    Both non-stationarities and consistencies are observed between subarrays. Specifically, characteristics such as azimuth angles, RMS AS, power of the LoS path, and Rician K factors vary across subarrays. Conversely, the envelope of dominant MPCs, RMS DS, angular distribution, and arrival time of the LoS paths show similarities, supporting effective distributed processing strategies. 

\subsubsection{Wideband Characteristics}
    Spatial channel characteristics across different sub-bands demonstrate similarities. The APAS and RMS AS derived from various sub-bands are comparable, enabling the utilization of spatial CSI on a sub-band basis to reduce computational complexity.

\subsubsection{Far-field Approximation of Near-field Non-stationary Channel} The near-field non-stationary ELAA channel can be effectively approximated as far-field subarray-wise under the U6G frequency band. This approximation is supported by the spread of the near-field channel towards multiple angular directions, validated by DFT analysis in both delay and angular domains. The NMSE of this approximation is about -20 dB and improves with an increase in the number of subarrays. These findings facilitate the adoption of efficient hardware structures and transmission strategies typically used in traditional massive MIMO systems through distributed processing.      

\subsection{Benefits of Distributed Processing}
Based on the characteristics discussed earlier, utilizing low-cost distributed processing structures and efficient algorithms at the subarray or sub-band level is an effective strategy to counter non-stationarities and leverage consistencies within U6G ELAA systems. This approach enables the system to adapt dynamically to variable channel conditions while optimizing resource utilization.

Therefore, we outline examples of distributed ELAA system configurations that are closely aligned with the measurement results detailed in this paper. These examples underscore the practical applications and benefits of distributed processing.

\subsubsection{Hardware Structure}
In massive MIMO systems, efficient hardware structures designed for far-field conditions have been proposed, such as those using DFT-based hybrid beamforming \cite{DFTTan,DFTHan}. In these systems, analog beamforming is executed using DFT matrices and implemented with low-cost hardware components, efficiently addressing far-field channel conditions. With the accurate subarray-wise far-field approximation confirmed by our measurements, the channel of a subarray can be effectively managed using a DFT-based structure, while the overall ELAA channel is processed through the integration of multiple subarrays. This method enhances system flexibility and scalability.

\subsubsection{Channel Estimation}

The consistencies observed between multiple subarrays and sub-bands suggest that similar spatial CSI can be shared among different subarrays and frequency bands. Consequently, CSI acquisition can be efficiently performed at the subarray or sub-band level, with complete system CSI obtained through extrapolation. Leveraging these consistencies allows for an effective strategy to manage the complexities of near-field characteristics and dual-band properties \cite{Dualband} in a distributed manner. This approach reduces the computational burden and simplifies the implementation of channel estimation processes across the system.

\section{Conclusion}
\label{sec:conclusion}
The U6G ELAA is emerging as a promising facilitator for the advancement of future wireless communication systems, supported by ongoing standardization efforts. Facing the uncertain channel characteristics of U6G ELAA systems, a fully digital U6G channel sounder was initially established, and measurement campaigns were conducted in typical scenarios to reveal the small-scale fading characteristics. Subsequently, the U6G ELAA channel was measured using a virtual array mechanism.

Based on these measurement results, we have investigated and modeled the small-scale fading characteristics of the U6G channel, including APAS, RMS AS, PDP, RMS DS, and the Rician K factor. We also compared these characteristics across various scenarios. It was found that sparse scattering characteristics are exhibited in different scenarios under the LoS condition, while the MPC characteristics fit the S-V model well in the delay domain. 
Considering the demands for low complexity and reduced hardware costs, the U6G ELAA channel characteristics were further analyzed from a distributed processing perspective. We explored the non-stationarities and consistencies of the aforementioned small-scale characteristics between subarrays and sub-bands. Additionally, the subarray-wise far-field approximation was verified to be accurate.
The findings from this study provide valuable measurement references for future system deployments and the design of algorithms for distributed U6G ELAA systems. These insights facilitate the development of advanced wireless technologies that leverage the unique properties of the U6G spectrum to enhance communication reliability and efficiency.

\vfill

\end{document}